\documentclass[aps,prd,preprint,superscriptaddress,showpacs]{revtex4}

\usepackage{graphicx}
\usepackage{ulem}
\usepackage{color}
\definecolor{My_red}        {cmyk}{0.00,1.00,1.00,0.20}

\def\nnb{\nonumber}

\def\bwt{\begin{widetext}}
\def\ewt{\end{widetext}}
\def\be{\begin{equation}}
\def\ee{\end{equation}}
\def\bea{\begin{eqnarray}}
\def\eea{\end{eqnarray}}
\def\bean{\begin{eqnarray*}}
\def\eean{\end{eqnarray*}}
\def\bary{\begin{array}}
\def\eary{\end{array}}
\def\bit{\begin{itemize}}
\def\eit{\end{itemize}}

\def\su5u1{SU(5) \times U(1)}
\def\fsu5u1{SU(5) \times U(1)'}
\def\so10{SO(10)}
\def\sq20{SO(10) \times SO(10)}

\def\nnb{\nonumber}

\def\bwt{\begin{widetext}}
\def\ewt{\end{widetext}}
\def\be{\begin{equation}}
\def\ee{\end{equation}}
\def\bea{\begin{eqnarray}}
\def\eea{\end{eqnarray}}
\def\bean{\begin{eqnarray*}}
\def\eean{\end{eqnarray*}}
\def\bary{\begin{array}}
\def\eary{\end{array}}
\def\bit{\begin{itemize}}
\def\eit{\end{itemize}}

\def\su5u1{SU(5) \times U(1)}
\def\fsu5u1{SU(5) \times U(1)'}
\def\so10{SO(10)}
\def\sq20{SO(10) \times SO(10)}

\def\bwt{\begin{widetext}}
\def\ewt{\end{widetext}}
\def\be{\begin{equation}}
\def\ee{\end{equation}}
\def\bea{\begin{eqnarray}}
\def\eea{\end{eqnarray}}
\def\bean{\begin{eqnarray*}}
\def\eean{\end{eqnarray*}}
\def\bary{\begin{array}}
\def\eary{\end{array}}
\def\bit{\begin{itemize}}
\def\eit{\end{itemize}}

\def\su5u1{SU(5) \times U(1)}
\def\fsu5u1{SU(5) \times U(1)'}
\def\so10{SO(10)}
\def\sq20{SO(10) \times SO(10)}

\begin{document}

\title{The Lightest CP-Even Higgs Boson Mass in the Testable 
Flipped $SU(5)\times U(1)_X$ Models from F-Theory}

\author{Yunjie Huo}

\affiliation{Key Laboratory of Frontiers in Theoretical Physics,
      Institute of Theoretical Physics, Chinese Academy of Sciences,
Beijing 100190, P. R. China }

\author{Tianjun Li}

\affiliation{Key Laboratory of Frontiers in Theoretical Physics,
      Institute of Theoretical Physics, Chinese Academy of Sciences,
Beijing 100190, P. R. China }

\affiliation{George P. and Cynthia W. Mitchell Institute for
Fundamental Physics, Texas A$\&$M University, College Station, TX
77843, USA }

\author{Dimitri V. Nanopoulos}

\affiliation{George P. and Cynthia W. Mitchell Institute for
Fundamental Physics,
 Texas A$\&$M University, College Station, TX 77843, USA }

\affiliation{Astroparticle Physics Group,
Houston Advanced Research Center (HARC),
Mitchell Campus, Woodlands, TX 77381, USA}

\affiliation{Academy of Athens, Division of Natural Sciences,
 28 Panepistimiou Avenue, Athens 10679, Greece }

\author{Chunli Tong}

\affiliation{Key Laboratory of Frontiers in Theoretical Physics,
      Institute of Theoretical Physics, Chinese Academy of Sciences,
Beijing 100190, P. R. China }

\date{\today}

\begin{abstract}

We study the lightest CP-even Higgs boson mass in
five kinds of testable flipped $SU(5)\times U(1)_X$
models from F-theory. Two kinds of models have 
vector-like particles around the TeV scale, while the 
other three kinds also have the vector-like 
particles at the intermediate scale that can be 
considered as messenger fields in gauge mediated 
supersymmetry breaking. We require that the Yukawa 
couplings for the TeV-scale vector-like particles and 
the third family of the Standard Model (SM) 
fermions are smaller than three from the 
electroweak scale to the $SU(3)_C\times SU(2)_L$ 
unification scale. With the 
two-loop renormalization group equation running 
for the gauge couplings and Yukawa couplings,
we obtain the maximal Yukawa couplings between
the TeV-scale vector-like particles and Higgs fields.
To calculate the lightest CP-even Higgs boson mass 
upper bounds, we employ the renormalization group 
 improved effective Higgs potential approach, 
and consider the two-loop leading contributions
in the supersymmetric SM and one-loop 
contributions from the TeV-scale vector-like particles.
We assume maximal mixings between the stops 
and between the TeV-scale vector-like scalars. The numerical 
results for these five kinds of models are roughly 
the same. In particular, we show that the lightest CP-even 
Higgs boson can have mass up to 146 GeV naturally, which 
is the current upper bound from the CMS and ATLAS collaborations.

\end{abstract}


\preprint{ACT-13-11, MIFPA-11-42}

\maketitle



\section{Introduction}

The Higgs boson mass in the Standard
Model (SM) is not stable against qunatum corrections
and has quadratic divergences. Because the reduced Planck
scale is about 16 order larger than the electroweak (EW) scale,
there exists huge fine-tuning to have the EW-scale 
Higgs boson mass, which is called the gauge hierarchy problem.
Supersymmetry is a symmetry between the bosonic and fermionic
states, and it naturally solves this
problem due to the cancellations between the bosonic and 
fermionic quantum corrections. 

In the Minimal Supersymmetric Standard Model (MSSM) with 
$R$ parity under which the SM particles are even 
while the supersymmetric particles (sparticles)
are odd, the $SU(3)_C\times SU(2)_L\times U(1)_Y$ gauge 
couplings can be unified around $2\times 10^{16}$ 
GeV~\cite{Langacker:1991an}, the lightest supersymmetric 
particle (LSP) such as the 
 neutralino can be a cold dark matter 
candidate~\cite{Ellis:1983ew, Goldberg:1983nd},
and the EW precision constraints can be
evaded, etc. Especially, the gauge coupling unification strongly
suggests Grand Unified Theories (GUTs), which can explain 
the SM fermion quantum numbers. 
However, in the supersymmetric $SU(5)$ models, 
there exist the doublet-triplet splitting problem and 
dimension-five proton decay problem. Interestingly, these problems
can be solved elegantly in the flipped $SU(5)\times U(1)_X$
models via missing partner mechanism~\cite{smbarr, dimitri, AEHN-0}.
Previously, the flipped $SU(5)\times U(1)_X$ models have been
constructed systematically in the free fermionic string 
constructions  at the Kac-Moody level one~\cite{Antoniadis:1988tt, Lopez:1992kg}. 
To solve
the little hierarchy problem between the traditional unification scale
and the string scale, two of us (TL and DVN) with Jiang have proposed the
testable flipped $SU(5)\times U(1)_X$ models, where the
TeV-scale vector-like particles are introduced~\cite{Jiang:2006hf}.
There is a two-step unifcation: the $SU(3)_C\times SU(2)_L$
gauge couplings are unified at the scale $M_{32}$ around
the usual GUT scale, and the $SU(5)\times U(1)_X$ gauge
couplings are unified at the final unification scale $M_{\cal{F}}$ 
around $5\times 10^{17}$ GeV~\cite{Jiang:2006hf}. Moreover,
such kind of models have been constructed locally  from
the F-theory model building~\cite{Beasley:2008dc, Jiang:2009zza}, 
and are dubbed as ${\cal F}$-$SU(5)$~\cite{Jiang:2009zza}. 
In particular, these models are very 
interesting from the phenomenological point of view~\cite{Jiang:2009zza}:
the vector-like particles can be observed at the Large Hadron Collider (LHC), 
 proton decay is within the reach of the future 
Hyper-Kamiokande~\cite{Nakamura:2003hk} and
Deep Underground Science and Engineering 
Laboratory (DUSEL)~\cite{DUSEL} experiments~\cite{Li:2009fq, Li:2010dp},
the hybrid inflation can be naturally realized, and the 
correct cosmic primodial density fluctuations can be 
generated~\cite{Kyae:2005nv}.  

With No-Scale boundary conditions 
at $M_{\cal{F}}$~\cite{Cremmer:1983bf}, 
two of us (TL and DVN) with Maxin and Walker have
 described an extraordinarily constrained ``golden point''~\cite{Li:2010ws} 
and ``golden strip''~\cite{Li:2010mi} that satisfied all the latest 
experimental constraints and has an imminently observable proton 
decay rate~\cite{Li:2009fq}. Especially, the UV boundary condition $B_{\mu}=0$
gives very strong constraint on the viable parameter space, where $B_{\mu}$
is the soft bilinear Higgs mass term in the MSSM.
In addition, exploiting a ``Super-No-Scale'' 
condition, we dynamically determined the universal gaugino mass
$M_{1/2}$ and the ratio of the Higgs Vacuum Expectation 
Values (VEVs)  $\tan\beta$.
Since $M_{1/2}$ is related to the modulus field of
the internal space in string models, we stabilized the modulus 
dynamically~\cite{Li:2010uu, Li:2011dw}. Interestingly, 
 the supersymmetric particle (sparticle) spectra 
generially have  a light stop and gluino, which are lighter than all the 
other squarks.
Thus, we can test such kinds of models at the LHC 
by looking for the ultra high jet signals~\cite{Li:2011hr, Maxin:2011hy}.
Moreover, the complete
viable parameter space in no-scale $\cal{F}$-$SU(5)$ has been
studied by considering a set of ``bare minimal'' experimental
constaints~\cite{Li:2011xu}. For the other LHC and dark matter 
phenomenological studies,
see Refs.~\cite{Li:2011in, Li:2011gh, Li:2011rp}.

It is well known that one of main LHC goals is to detect the SM or SM-like
Higgs boson. Recently, both the CMS~\cite{CMSLP} and ATLAS~\cite{ATLASLP} collaborations
have presented their combined searches for the SM Higgs boson, base on
the integrated luminosities between $1~{\rm fb}^{-1}$ and $2.3~{\rm fb}^{-1}$
depending on the search channels. For the light SM Higgs boson mass region 
preferred by the EW precision data, they have excluded the SM Higgs boson 
with mass larger than 145 GeV and 146 GeV, respectively.  
In the no-scale $\cal{F}$-$SU(5)$, 
the lightest CP-even Higgs boson mass is generically about 120 GeV if the
contributions from the vector-like particles are neglected~\cite{LMNW-P}.
Thus, the interesting question is whether the lightest CP-even Higgs boson
can have mass up to 146 GeV naturally if we include the contributions from
the additional vector-like particles.

In this paper, we consider five kinds of testable 
flipped $SU(5)\times U(1)_X$ models from F-theoy.
Two kinds of models only have vector-like particles around the TeV scale.
Because the gauge mediated supersymmetry breaking can be realized naturally in
the F-theory GUTs~\cite{Heckman:2008qt},  
we also introduce vector-like particles
with mass around $10^{11}~{\rm GeV}$~\cite{Heckman:2008qt}, which can be considered
as messenger fields, in the other three kinds of models. We require that
the Yukawa couplings for the TeV-scale vector-like particles and 
the third family of the SM fermions are smaller than three from the EW scale to
the scale $M_{32}$ from the perturbative bound, {\it i.e.}, the
Yukawa coupling squares are less than $4\pi$.
 With the two-loop Renormalization Group 
Equation (RGE) running for the gauge couplings and Yukawa couplings,
we obtain the maximal Yukawa couplings for
the  TeV-scale vector-like particles. To calculate the lightest CP-even
Higgs boson mass upper bounds, we employ the Renormalization Group (RG) improved
effective Higgs potential approach, and consider the two-loop leading contributions
in the MSSM and one-loop contributions from the TeV-scale 
vector-like particles. For simplicity, we assume that the 
 mixings both between the stops and between  the TeV-scale vector-like scalars are
maximal. In general, we shall increase the lightest CP-even Higgs boson
mass upper bounds if we increase the supersymmetry breaking scale
or decrease the TeV-scale vector-like particle masses.
The numerical results for our five kinds of models are roughly the same. 
For the TeV-scale vector-like particles and sparticles with masses
 around 1~TeV, we show that  the lightest 
CP-even Higgs boson can have mass up to 146 GeV naturally.

This paper is organized as follows. In Section II, we briefly
review the testable flipped $SU(5)\times U(1)_X$ models from
F-theory and present five kinds of models. We calculate the
lightest CP-even Higgs boson mass upper bounds 
 in Section III. Section IV is our conclusion.
In Appendices, we present all the RGEs in five kinds of models.



\section{Testable Flipped $SU(5)\times U(1)_X$ Models from F-Theory}

We first briefly review the minimal flipped
$SU(5)$ model~\cite{smbarr, dimitri, AEHN-0}.
The gauge group for flipped $SU(5)$ model is
$SU(5)\times U(1)_{X}$, which can be embedded into $SO(10)$ model.
We define the generator $U(1)_{Y'}$ in $SU(5)$ as
\bea
T_{\rm U(1)_{Y'}}={\rm diag} \left(-{1\over 3}, -{1\over 3}, -{1\over 3},
 {1\over 2},  {1\over 2} \right).
\label{u1yp}
\eea
The hypercharge is given by
\bea
Q_{Y} = {1\over 5} \left( Q_{X}-Q_{Y'} \right).
\label{ycharge}
\eea

There are three families of the SM fermions
whose quantum numbers under $SU(5)\times U(1)_{X}$ are
\bea
F_i={\mathbf{(10, 1)}},~ {\bar f}_i={\mathbf{(\bar 5, -3)}},~
{\bar l}_i={\mathbf{(1, 5)}},
\label{smfermions}
\eea
where $i=1, 2, 3$. The SM particle assignments in $F_i$, ${\bar f}_i$
and ${\bar l}_i$ are
\bea
F_i=(Q_i, D^c_i, N^c_i),~{\overline f}_i=(U^c_i, L_i),~{\overline l}_i=E^c_i~,~
\label{smparticles}
\eea
where $Q_i$ and $L_i$ are respectively the superfields of the left-handed
quark and lepton doublets, $U^c_i$, $D^c_i$, $E^c_i$ and $N^c_i$ are the
$CP$ conjugated superfields for the right-handed up-type quarks,
down-type quarks, leptons and neutrinos, respectively.
To generate the heavy right-handed neutrino masses, we introduce
three SM singlets $\phi_i$~\cite{Georgi:1979dq}.

To break the GUT and electroweak gauge symmetries, we introduce two pairs
of Higgs representations
\bea
H={\mathbf{(10, 1)}},~{\overline{H}}={\mathbf{({\overline{10}}, -1)}},
~h={\mathbf{(5, -2)}},~{\overline h}={\mathbf{({\bar {5}}, 2)}}.
\label{Higgse1}
\eea
We label the states in the $H$ multiplet by the same symbols as in
the $F$ multiplet, and for ${\overline H}$ we just add ``bar'' above the fields.
Explicitly, the Higgs particles are
\bea
H=(Q_H, D_H^c, N_H^c)~,~
{\overline{H}}= ({\overline{Q}}_{\overline{H}}, {\overline{D}}^c_{\overline{H}},
{\overline {N}}^c_{\overline H})~,~\,
\label{Higgse2}
\eea
\bea
h=(D_h, D_h, D_h, H_d)~,~
{\overline h}=({\overline {D}}_{\overline h}, {\overline {D}}_{\overline h},
{\overline {D}}_{\overline h}, H_u)~,~\,
\label{Higgse3}
\eea
where $H_d$ and $H_u$ are one pair of Higgs doublets in the MSSM.
We also add one SM singlet $\Phi$.

To break the $SU(5)\times U(1)_{X}$ gauge symmetry down to the SM
gauge symmetry, we introduce the following Higgs superpotential at the GUT scale
\bea
{\it W}_{\rm GUT}=\lambda_1 H H h + \lambda_2 {\overline H} {\overline H} {\overline
h} + \Phi ({\overline H} H-M_{\rm H}^2)~.~
\label{spgut}
\eea
There is only
one F-flat and D-flat direction, which can always be rotated along
the $N^c_H$ and ${\overline {N}}^c_{\overline H}$ directions. So, we obtain that
$<N^c_H>=<{\overline {N}}^c_{\overline H}>=M_{\rm H}$. In addition, the
superfields $H$ and ${\overline H}$ are eaten and acquire large masses via
the supersymmetric Higgs mechanism, except for $D_H^c$ and
${\overline {D}}^c_{\overline H}$. The superpotential $ \lambda_1 H H h$ and
$ \lambda_2 {\overline H} {\overline H} {\overline h}$ couple the $D_H^c$ and
${\overline {D}}^c_{\overline H}$ with the $D_h$ and ${\overline {D}}_{\overline h}$,
respectively, to form the massive eigenstates with masses
$2 \lambda_1 <N_H^c>$ and $2 \lambda_2 <{\overline {N}}^c_{\overline H}>$. So, we
naturally have the doublet-triplet splitting due to the missing
partner mechanism~\cite{AEHN-0}.
Because the triplets in $h$ and ${\overline h}$ only have
small mixing through the $\mu$ term, the Higgsino-exchange mediated
proton decay are negligible, {\it i.e.},
we do not have the dimension-5 proton
decay problem.

The SM fermion masses are from the following
superpotential
\bea
{ W}_{\rm Yukawa} = y_{ij}^{D}
F_i F_j h + y_{ij}^{U \nu} F_i  {\overline f}_j {\overline
h}+ y_{ij}^{E} {\overline l}_i  {\overline f}_j h + \mu h {\overline h}
+ y_{ij}^{N} \phi_i {\overline H} F_j~,~\,
\label{potgut}
\eea
where $y_{ij}^{D}$, $y_{ij}^{U \nu}$, $y_{ij}^{E}$ and $y_{ij}^{N}$
are Yukawa couplings, and $\mu$ is the bilinear Higgs mass term.

After the $SU(5)\times U(1)_X$ gauge symmetry is broken down to the SM gauge
symmetry, the above superpotential gives
\bea
{ W_{SSM}}&=&
y_{ij}^{D} D^c_i Q_j H_d+ y_{ji}^{U \nu} U^c_i Q_j H_u
+ y_{ij}^{E} E^c_i L_j H_d+  y_{ij}^{U \nu} N^c_i L_j H_u \nnb \\
&& +  \mu H_d H_u+ y_{ij}^{N}
\langle {\overline {N}}^c_{\overline H} \rangle \phi_i N^c_j
 + \cdots (\textrm{decoupled below $M_{GUT}$}).
\label{poten1}
\eea

Similar to the flipped $SU(5)\times U(1)_X$ models
with string-scale gauge coupling
unification~\cite{Jiang:2006hf, Lopez:1995cs},
we introduce vector-like particles which form  complete
flipped $SU(5)\times U(1)_X$ multiplets.
The quantum numbers for these additional vector-like particles
 under the $SU(5)\times U(1)_X$ gauge symmetry are
\begin{eqnarray}
&& XF ={\mathbf{(10, 1)}}~,~{\overline{XF}}={\mathbf{({\overline{10}}, -1)}}~,~\\
&& Xf={\mathbf{(5, 3)}}~,~{\overline{Xf}}={\mathbf{({\overline{5}}, -3)}}~,~\\
&& Xl={\mathbf{(1, -5)}}~,~{\overline{Xl}}={\mathbf{(1, 5)}}~,~\\
&& Xh={\mathbf{(5, -2)}}~,~{\overline{Xh}}={\mathbf{({\overline{5}}, 2)}}~,~ \\
&& XT ={\mathbf{(10, -4)}}~,~{\overline{XT}}={\mathbf{({\overline{10}}, 4)}}~.~\,
\end{eqnarray}

Moreover,  the particle contents from the decompositions of
$XF$, ${\overline{XF}}$, $Xf$, ${\overline{Xf}}$,
$Xl$, ${\overline{Xl}}$, $Xh$, ${\overline{Xh}}$,
$XT$, and ${\overline{XT}}$, under the SM gauge
symmetry are
\begin{eqnarray}
&& XF = (XQ, XD^c, XN^c)~,~ {\overline{XF}}=(XQ^c, XD, XN)~,~\\
&& Xf=(XU, XL^c)~,~ {\overline{Xf}}= (XU^c, XL)~,~\\
&& Xl= XE~,~ {\overline{Xl}}= XE^c~,~\\
&& Xh=(XD, XL)~,~ {\overline{Xh}}= (XD^c, XL^c)~,~\\
&& XT = (XY, XU^c, XE)~,~ {\overline{XT}}=(XY^c, XU, XE^c)~.~\,
\end{eqnarray}
Under the $SU(3)_C \times SU(2)_L \times U(1)_Y$ gauge
symmetry, the quantum numbers for the extra vector-like
particles are
\begin{eqnarray}
&& XQ={\mathbf{(3, 2, {1\over 6})}}~,~
XQ^c={\mathbf{({\bar 3}, 2,-{1\over 6})}} ~,~\\
&& XU={\mathbf{({3},1, {2\over 3})}}~,~
XU^c={\mathbf{({\bar 3},  1, -{2\over 3})}}~,~\\
&& XD={\mathbf{({3},1, -{1\over 3})}}~,~
XD^c={\mathbf{({\bar 3},  1, {1\over 3})}}~,~\\
&& XL={\mathbf{({1},  2,-{1\over 2})}}~,~
XL^c={\mathbf{(1,  2, {1\over 2})}}~,~\\
&& XE={\mathbf{({1},  1, {-1})}}~,~
XE^c={\mathbf{({1},  1, {1})}}~,~\\
&& XN={\mathbf{({1},  1, {0})}}~,~
XN^c={\mathbf{({1},  1, {0})}}~,~\\
&& XY={\mathbf{({3}, 2, -{5\over 6})}}~,~
XY^c={\mathbf{({\bar 3}, 2, {5\over 6})}} ~.~\
\end{eqnarray}

To separate the mass scales $M_{32}$ and $M_{\cal F}$ in our F-theory
flipped $SU(5)\times U(1)_X$ models,
we need to introduce sets of vector-like particles
around the TeV scale or intermediate scale whose contributions to the one-loop
beta functions satisfy $\Delta b_1 < \Delta b_2 = \Delta b_3$.
To avoid the Landau pole problem, we have shown that there are
only five possible such sets of vector-like
 particles as follows
due to the quantizations of the one-loop beta functions~\cite{Jiang:2006hf}
\begin{eqnarray}
&& Z0: XF+{\overline{XF}}~;~\\
&& Z1: XF+{\overline{XF}}+Xl+{\overline{Xl}} ~;~\\
&&  Z2: XF+{\overline{XF}}+Xf+{\overline{Xf}} ~;~\\
&&  Z3: XF+{\overline{XF}} + Xl+{\overline{Xl}}
+Xh+{\overline{Xh}}  ~;~\\
&&  Z4: XF+{\overline{XF}}+Xh+{\overline{Xh}}~.~\,
\end{eqnarray}

We have systematically constructed  flipped $SU(5)\times U(1)_X$ models with
generic sets of vector-like particles around the TeV scale and/or
around the intermediate scale from the F-theory. In addition,
 gauge mediated supersymmetry breaking can be realized naturally in
the F-theory GUTs~\cite{Heckman:2008qt}, and there may exist vector-like particles as
the messenger fields at the intermediate scale around $10^{11}$~GeV~\cite{Heckman:2008qt}.
Therefore, in this paper, we shall calculate the lightest CP-even Higgs boson 
mass in five kinds of the flipped $SU(5)\times U(1)_X$ models
from F-theory: (i) In Model I, we introduce the $Z0$ set of 
vector-like particles $(XF, ~{\overline{XF}})$ at the TeV scale,
and we shall add superheavy vector-like particles around $M_{32}$ so
that the $SU(5)\times U(1)_X$ unification scale is smaller than
the reduced Planck scale; (ii) In Model II, we introduce the
vector-like particles $(XF, ~{\overline{XF}})$ at the TeV scale
and the vector-like particles $(Xf, ~{\overline{Xf}})$  at the 
 intermediate scale  which can be considered as the messenger fields;
(iii) In Model III, we introduce the
vector-like particles $(XF, ~{\overline{XF}})$ at the TeV scale
and the vector-like particles  $(Xf, ~{\overline{Xf}})$ 
and  $(Xl, ~{\overline{Xl}})$ at the 
 intermediate scale; (iv) 
 In Model IV, we introduce  the $Z1$ set of 
vector-like particles $(XF, ~{\overline{XF}})$ and
 $(Xl, ~{\overline{Xl}})$ at the TeV scale;
(v) In Model V, we introduce the
vector-like particles $(XF, ~{\overline{XF}})$ and
  $(Xl, ~{\overline{Xl}})$ at the TeV scale, and
 the vector-like particles $(Xf, ~{\overline{Xf}})$  at the 
 intermediate scale.
In particular, we emphasize that 
the vector-like particles at the intermediate scale in Models II, III, and V
will give us the generalized  gauge medidated supersymmetry breaking
if they are the messenger fields~\cite{Li:2010hi}. By the way,
if we introduce the vector-like particles $(Xh, ~{\overline{Xh}})$ at
the intermediate scale which are the traditional messenger fields in gauge mediation,
the discussions are similar and the numerical results are almost the same.
Thus, we will not study such kind of models here.

For simplicity, we assume that the masses for the vector-like particles around 
the TeV scale or the intermediate scale are universal. Also, we denote the 
universal mass for the vector-like particles at the TeV
scale as $M_V$, and the universal mass for the  vector-like particles at the 
intermediate scale as $M_I$. With this convention, we present the 
vector-like particle contents
of our five kinds of models in Table~\ref{Model-PC}.
In the following discussions, we shall choose $M_{I}=1.0\times 10^{11}$~GeV. 
Moreover, we will assume 
universal supersymmetry breaking at low energy and denote
the universal supersymmetry breaking scale as $M_S$.

\begin{table}[htb]
\begin{center}
\begin{tabular}{|c|c|c|}
\hline
Models &  Vector-Like Particles at $M_V$    & Vector-Like Particles at  $M_{I}$   \\
\hline
\hline
~Model I~& ~ ($XF$, $\overline{XF}$)
&  \\
\hline
~Model II~& ~ ($XF$, $\overline{XF}$)
& ~($Xf$,  $\overline{Xf}$) ~ \\
\hline
~Model III~& ~ ($XF$, $\overline{XF}$)
& ~($Xf$,  $\overline{Xf}$),~ ($Xl$, $\overline{Xl}$)\\
\hline
~Model IV~& ~ ($XF$, $\overline{XF}$),~ ($Xl$, $\overline{Xl}$)~
& \\
\hline
~Model V~& ~ ($XF$, $\overline{XF}$),~ ($Xl$, $\overline{Xl}$)~
& ~($Xf$,  $\overline{Xf}$) ~ \\
\hline
\end{tabular}
\end{center}
\caption{The vector-like particle contents in Model I, Model II,  Model III,
Model IV, and Model V. }
\label{Model-PC}
\end{table}

 It is well known that there exists a few pecent fine-tuning for
the lightest CP-even Higgs boson mass in the MSSM to be 
larger than 114.4 GeV. In all the above five kinds of models,
we have the vector-like particles $XF$ and $\overline{XF}$ at
the TeV scale. Then we can introduce the
 following Yukawa interaction terms between  the MSSM Higgs fields
and these vector-like particles in the superpotential in the
flipped $SU(5)\times U(1)_X$ models:
\begin{eqnarray}
W &=& {1\over2} Y_{xd} XF XF h + {1\over2}
Y_{xu} \overline{XF} \overline{XF} \overline{h}~,~\,
\end{eqnarray}
where $Y_{xd}$ and $Y_{xu}$ are Yukawa couplings.
After the gauge symmetry $SU(5)\times U(1)_X$ is broken down
to the SM gauge symmetry, we have the following relevant
Yukawa coupling terms in the superpotential
\begin{eqnarray}
W &=&  Y_{xd} XQ XD^c H_d +
 Y_{xu} XQ^c XD H_u ~.~\,
\end{eqnarray}

To have the upper bounds on the lightest CP-even
Higgs boson mass, we first need to calculate the
upper bounds on the Yukawa couplings  $Y_{xu}$ and $Y_{xd}$.
 In this paper,  employing the two-loop RGE running,
we will require that all
the Yukawa couplings including $Y_{xu}$ and $Y_{xd}$ are smaller than three
(perturbative bound) 
below the $SU(3)_C \times SU(2)_L$ unification
scale $M_{32}$ for simplicity since $M_{32}$ is close to 
the $SU(5) \times U(1)_X$ unification scale $M_{\cal F}$. 
The other point is that above the scale $M_{32}$, 
there might exist other superheavy threshold corrections 
and then the RGE running for the gauge couplings and Yukawa couplings 
 might be very complicated. Moreover, we will
not give the two-loop RGEs in the SM and 
the MSSM, which can be easily found in the literatures, for example,
in the Refs.~\cite{Barger:1992ac, Martin:1993zk}.
We shall present the RGEs in the SM with vector-like particles, and 
Models I to V in the Appendices A, B, C, 
D, E, and F, respectively.



\section{The Lightest CP-Even Higgs Boson Mass}

In our calculations, we employ the RG improved 
effective Higgs potential approach. The two-loop leading
contributions to the lightest CP-even Higgs boson mass $m_h$
in the MSSM are~\cite{Okada:1990vk,Carena:1995bx}
\begin{eqnarray}
[m_h^2]_{\mbox{MSSM}}&=&M_Z^2\cos^22\beta(1-\frac{3}{8\pi^2}\frac{m_t^2}{v^2}t)\nonumber\\
&&+\frac{3}{4\pi^2}\frac{m_t^4}{v^2}[t+\frac{1}{2}X_t
+\frac{1}{(4\pi)^2}(\frac{3}{2}\frac{m_t^2}{v^2}-32\pi\alpha_s)(X_tt+t^2)],
\end{eqnarray}
where $M_Z$ is the $Z$ boson mass, $m_t$ is the $\overline{MS}$ top
quark mass,  $v$ is the SM Higgs VEV,
and $\alpha_S$ is the strong coupling constant. Also, $t$ and $X_t$ are
given as follows
\begin{eqnarray}
t=\mbox{log}\frac{M_S^2}{M_t^2},~~~X_t=\frac{2{\tilde
A}_t^2}{M_S^2}(1-\frac{{\tilde A}_t^2}{12M_S^2}),~~~{\tilde A}_{t}=A_{t}-\mu\cot\beta,
\end{eqnarray}
where $M_t$ is the top quark pole mass,
and $A_t$ denotes the trilinear soft term for the top quark Yukawa coupling term.

Moreover, we use the RG-improved one-loop effective Higgs potential
approach to calculate the contributions to the lightest CP-even Higgs boson 
mass from the vector-like
particles~\cite{Babu:2008ge,Martin:2009bg}. Such contributions in our models are
\begin{eqnarray}
\Delta m_h^2&=&-\frac{N_c}{8\pi^2}M_Z^2\cos^22\beta({\hat
Y}_{xu}^2+{\hat Y}_{xd}^2)t_V+\frac{N_cv^2}{4\pi^2}\times\{{\hat
Y}_{xu}^4[t_V+\frac{1}{2}X_{xu}]\nonumber\\
&&+{\hat Y}_{xu}^3{\hat
Y}_{xd}[-\frac{2M_S^2(2M_S^2+M_V^2)}{3(M_S^2+M_V^2)^2}-\frac{{\tilde
A}_{xu}(2{\tilde A}_{xu}+{\tilde
A}_{xd})}{3(M_S^2+M_V^2)}]\nonumber\\
&&+{\hat Y}_{xu}^2{\hat
Y}_{xd}^2[-\frac{M_S^4}{(M_S^2+M_V^2)^2}-\frac{({\tilde
A}_{xu}+{\tilde A}_{xd})^2}{3(M_S^2+M_V^2)}]\nonumber\\
&&+{\hat Y}_{xu}{\hat
Y}_{xd}^3[-\frac{2M_S^2(2M_S^2+M_V^2)}{3(M_S^2+M_V^2)^2}-\frac{{\tilde
A}_{xd}(2{\tilde A}_{xd}+{\tilde A}_{xu})}{3(M_S^2+M_V^2)}]+{\hat
Y}_{xd}^4[t_V+\frac{1}{2}X_{xd}]\},\label{Delta mhs}
\end{eqnarray}
where
\begin{eqnarray}
&&{\hat Y}_{xu}=Y_{xu}\sin\beta,~~~~~{\hat
Y}_{xd}=Y_{xd}\cos\beta,~~~~~
t_V=\mbox{log}\frac{M_S^2+M_V^2}{M_V^2},\nonumber\\
&&X_{xu}=-\frac{2M_S^2(5M_S^2+4M_V^2)-4(3M_S^2+2M_V^2){\tilde
A}_{xu}^2+{\tilde A}_{xu}^4}{6(M_V^2+M_S^2)^2},\nonumber\\
&&X_{xd}=-\frac{2M_S^2(5M_S^2+4M_V^2)-4(3M_S^2+2M_V^2){\tilde
A}_{xd}^2+{\tilde A}_{xd}^4}{6(M_V^2+M_S^2)^2},\nonumber\\
&&{\tilde A}_{xu}=A_{xu}-\mu\cot\beta,~~~~~~~{\tilde
A}_{xd}=A_{xd}-\mu\tan\beta,
\end{eqnarray}
where ${A}_{xu}$ and  ${A}_{xd}$ denote the 
supersymmetry breaking trilinear soft terms for 
the superpotential Yukawa terms $Y_{xu} XQ^c XD H_u $
and  $Y_{xd} XQ XD^c H_d$, respectively.

The third, fourth, fifth, and sixth terms in Eq.~(\ref{Delta mhs})
are suppressed by the inverses of
$\tan\beta$, $\tan^2\beta$, $\tan^3\beta$, and
$\tan^4\beta$, respectively. To have the lightest CP-even Higgs boson mass
upper bounds, we usually need $\tan\beta \sim 22 $
 from the numerical calculations. 
 Especially, in order to increase the
 lightest CP-even Higgs boson mass, we should choose relatively large $Y_{xu}$ and
small $Y_{xd}$~\cite{Babu:2008ge,Martin:2009bg}. Thus, for simplicity, 
we only employ the first and second terms 
in our calculations, {\it i.e.}, the first line of Eq.~(\ref{Delta mhs}). 
 In order to have larger corrections to the lightest CP-even Higgs
boson mass, we consider the maximal mixings $X_t$ and
$X_{xu} $ respectively for both the stops and the TeV-scale vector-like
scalars, {\it i.e.}, $X_t=6$ with ${\tilde A}_t^2=6M_S^2$, and
$X_{xu}=\frac{8}{3}+\frac{M_S^2(5M_S^2+4M_V^2)}{3(M_S^2+M_V^2)}$
with ${\tilde A}_{xu}^2=6M_S^2+4M_V^2$.

\begin{figure}[h]
  \begin{center}
      \includegraphics[width=6in]{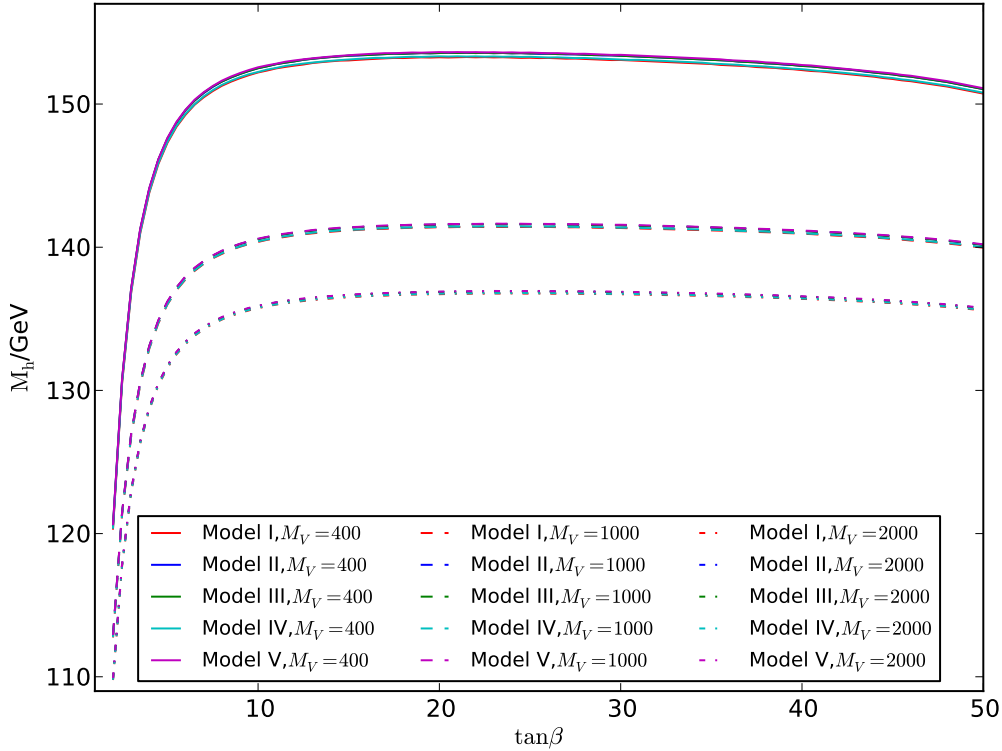}
  \end{center}
\vspace{-1cm}\caption{(color online). The upper bounds on the lightest
CP-even Higgs boson mass versus $\tan\beta$ for our five kinds of models with
$Y_{xd}=0$, $M_S=800$~GeV, and $M_I=1.0\times 10^{11}$~GeV. The upper lines,
middle lines, and lower lines are for  $M_V=400$~GeV, 1000~GeV,
and 2000~GeV, respectively. }
\label{varying-with-tanbeta-yxd=0}
\end{figure}

In this Section, we shall calculate the lightest CP-even Higgs boson mass in 
our five kinds of models.  The relevant parameters are the universal supersymmetry 
breaking scale $M_S$,  the  light vector-like particle mass $M_V$, the 
intermediate scale $M_I$, the 
mixing terms $X_t$ and $X_V$ respectively for the stops and TeV-scale vector-like
scalars, and the two new Yukawa couplings for TeV-scale vector-like particles
$Y_{xu}$ and $Y_{xd}$. 
Because we consider low energy supersymmetry,  we choose $M_S$ from 
360~GeV to 2~TeV. In order
to increase the lightest CP-even Higgs boson mass, we need to choose small $M_V$ as well.
The experimental lower bound on $M_V$ is about
325~GeV~\cite{Graham:2009gy}, so we will choose $M_V$ from  360~GeV to 2~TeV.
In our numerical calculations, we will use the SM input parameters at scale 
$M_Z$ from Particle Data Group~\cite{Nakamura:2010zzi}. In particular, we use the updated 
top quark pole mass $M_t=172.9$~GeV, and the corresponding 
$\overline{MS}$ top quark mass $m_t=163.645$~GeV~\cite{Nakamura:2010zzi}.

In this paper, we require that all the Yukawa couplings for both the TeV-scale
vector-like particles and the third family of SM fermions
 are less than three (perturbative bound) from the
EW scale to  the scale $M_{32}$. To obtain the upper bounds on the Yukawa couplings 
$Y_{xu}$ and $Y_{xd}$ at low energy, we consider the two-loop RGE running 
for both the SM gauge couplings and all the Yukawa couplings. The only exception is 
that when $M_V<M_S$, we use the two-loop RGE running 
for the SM gauge couplings and
one-loop RGE running for all the Yukawa couplings
 from $M_V$ to $M_S$, see Appendix A for details. Because in this case 
$M_V$ is still close
to $M_S$, such small effects are negligible. After we obtain the upper bounds on 
 $Y_{xu}$ and $Y_{xd}$,
we use the maximal $Y_{xu}$ to
calculate the upper bounds on the lightest CP-even Higgs boson mass  with the maximal
 mixings for  stops and TeV-scale
vector-like scalars.

\begin{figure}[t]
      \begin{center}
            \includegraphics[width=6in]{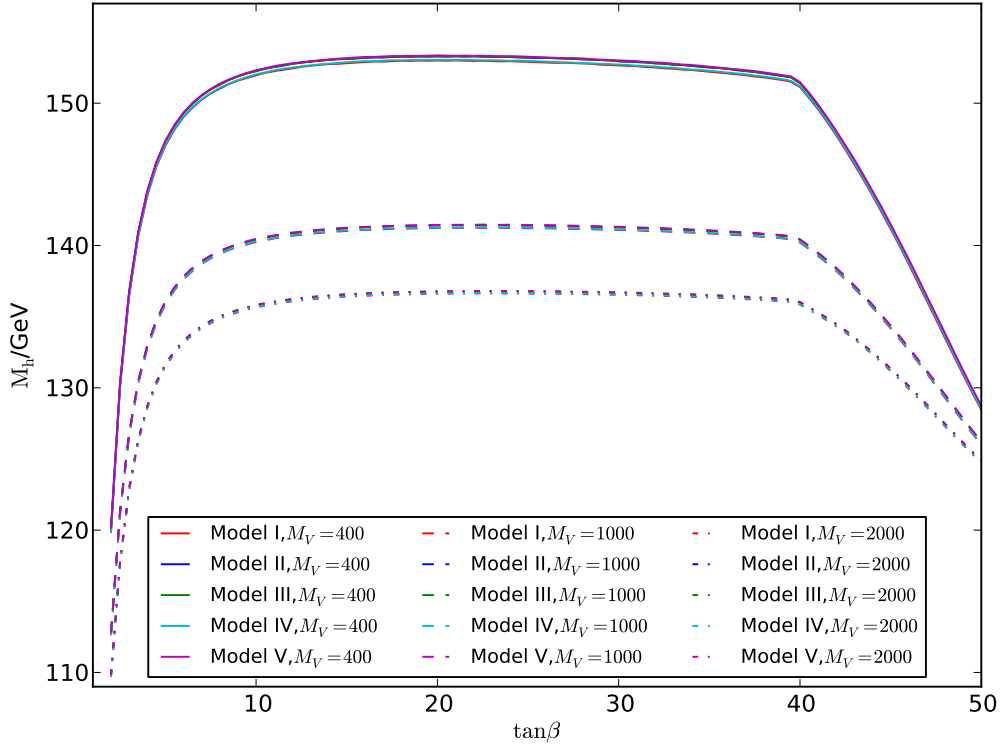}
      \end{center}
\vspace{-1cm}\caption{(color online). The upper bounds on the lightest
CP-even Higgs boson mass versus $\tan\beta$ for our five kinds of models with
$Y_{xd}(M_V)=Y_{xu}(M_V)$, $M_S=800$~GeV, and $M_I=1.0\times 10^{11}$~GeV. The upper lines,
middle lines, and lower lines are for  $M_V=400$~GeV, 1000~GeV,
and 2000~GeV, respectively. }
\label{varying-with-tanbeta-yxd=yxu}
\end{figure}

First, we consider $Y_{xd}=0$, $M_S=800$~GeV, and $M_I=1.0\times 10^{11}$~GeV. We choose
three values for $M_V$: $M_V=400$~GeV, 1000~GeV,
and 2000~GeV. In Fig.~\ref{varying-with-tanbeta-yxd=0}, we present the upper
bounds on the lightest CP-even Higgs boson mass by varying $\tan\beta$ from
2 to 50. We find that for the same $M_V$, the upper bounds on the lightest
CP-even Higgs boson mass are almost the same for five kinds of models.
In particular, the small differences are less than 0.4 GeV. Because the
gauge couplings will give negative contributions to the Yukawa coupling
RGEs, we will have a little bit larger maximal Yukawa couplings $Y_{xu}$ if
the vector-like particles contribute more to the gauge coupling
RGE running. Thus, the Model order for the lightest CP-even Higgs boson
mass upper bounds from small to large is Model I, Model IV, Model II, Model III,
Model V. Also, the upper bounds on the lightest CP-even Higgs boson
mass will decrease when we increase $M_V$, which is easy
to understand from physics point of view. Moreover,
the maximal Yukawa couplings $Y_{xu}$ are about 0.96, 1.03, and 1.0
for $\tan\beta=2$, $\tan\beta \sim 23$,  and $\tan\beta=50$, respectively.
 In addition, for $M_V=400$~GeV and $\tan\beta \simeq 21$, 
$M_V=1000$~GeV and $\tan\beta \simeq 23.5$, and $M_V=2000$~GeV and $\tan\beta \simeq 24.5$,
 we obtain the lightest CP-even Higgs boson mass upper bounds
around 153.5~GeV, 141.6~GeV, and 136.8~GeV, respectively.

\begin{figure}[t]
      \begin{center}
            \includegraphics[width=6in]{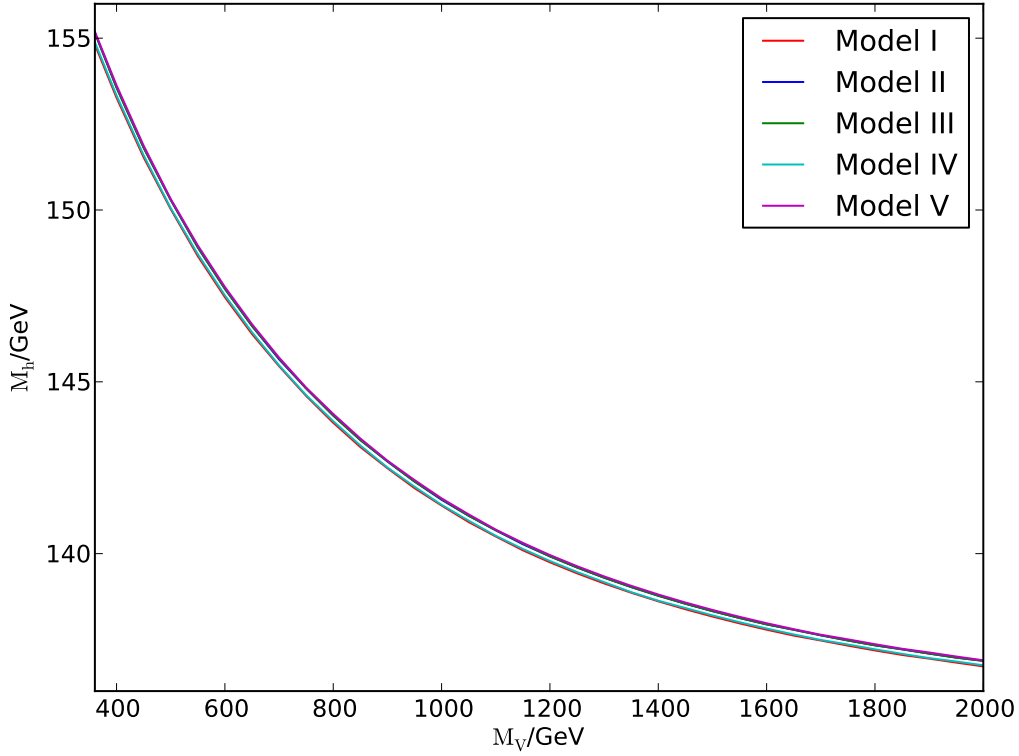}
      \end{center}
\vspace{-1cm}\caption{(color online). The upper bounds on the lightest
CP-even Higgs boson mass versus $M_V$ for  our five kinds of models with
 $Y_{xd}=0$, $\tan\beta=20$, $M_S=800$~GeV, and  $M_I=1.0\times 10^{11}$~GeV. }
\label{varying-with-M_V-yxd=0}
\end{figure}

Second, we consider $Y_{xd}=Y_{xu}$ at the scale $M_V$,
 $M_S=800$~GeV, and $M_I=1.0\times 10^{11}$~GeV. We choose
three values for $M_V$: $M_V=400$~GeV, 1000~GeV,
and 2000~GeV. In Fig.~\ref{varying-with-tanbeta-yxd=yxu}, we present the upper
bounds on the lightest CP-even Higgs boson mass by varying $\tan\beta$ from
2 to 50.  For $\tan\beta < 40$, we find that the lightest CP-even Higgs boson mass 
upper bounds are almost the same as those in Fig.~\ref{varying-with-tanbeta-yxd=0}. 
However, for $\tan\beta > 40$, we find that the lightest CP-even Higgs boson mass 
upper bounds decrease fast when $\tan\beta$ increases. At $\tan\beta=50$,
the upper bounds on the lightest CP-even Higgs boson mass are smaller than
130 GeV for all our scenarios. The reasons are the following: 
for $\tan\beta<40$, the Yukawa couplings 
$Y_{xu}$ and $Y_{t}$ are easy to  run out of the perturbative bound, while 
for $\tan\beta>40$, the Yukawa couplings $Y_{xd}$, $Y_{b}$,
and especially $Y_{\tau}$ are easy to run out, where
$Y_t$, $Y_{b}$ and $Y_{\tau}$ are Yukawa couplings for the top quark,
bottom quark, and tau lepton, respectively. In particular, for $\tan\beta =50$,
the maximal Yukawa couplings $Y_{xd}=Y_{xu}$ are as small as 0.67
while they are about 1.025 for $\tan\beta<40$.

\begin{figure}[t]
      \begin{center}
            \includegraphics[width=6in]{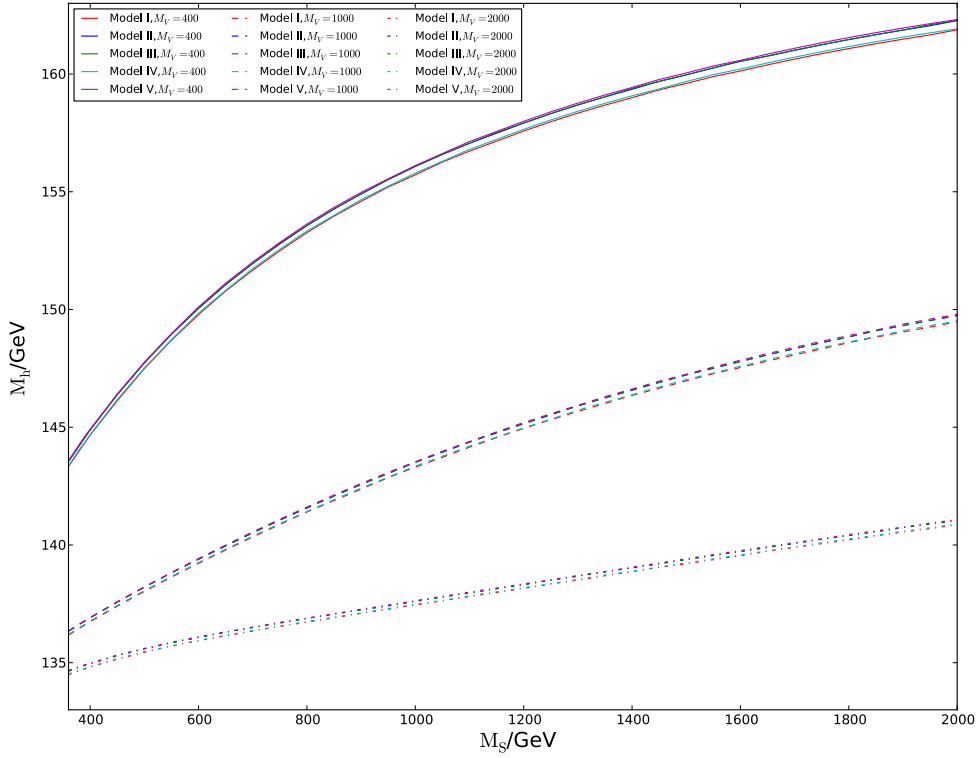}
      \end{center}
\vspace{-1cm}\caption{(color online). The upper bounds on the lightest
CP-even Higgs boson mass versus $M_S$ for our five kinds of models
  with $Y_{xd}=0$,  $\tan\beta=20$, and $M_I=1.0\times 10^{11}$~GeV. The upper lines,
middle lines, and lower lines are for  $M_V=400$~GeV, 1000~GeV,
and 2000~GeV, respectively.}
\label{varying-with-M_S-yxd=0}
\end{figure}

Third, we consider $Y_{xd}=0$, $\tan\beta=20$,  $M_S=800$~GeV, and  $M_I=1.0\times 10^{11}$~GeV.
In Fig.~\ref{varying-with-M_V-yxd=0}, we present the upper bounds on
the lightest CP-even Higgs boson mass by varying $M_V$ from
360~GeV to 2~TeV. We can see that as
the value of $M_V$ increases from 360~GeV to 2~TeV, the upper bounds on the lightest 
CP-even Higgs boson
mass decrease from 155~GeV to 137~GeV. In particular, to have the lightest 
CP-even Higgs boson mass
upper bounds larger than 146 GeV, we obtain that $M_V$ is smaller than about 700 GeV.
 Moreover, the maximal Yukawa couplings $Y_{xu}$ vary
only a little bit, decreasing from about 1.029 to 1.016 for $M_V$ 
from 360~GeV to 2~TeV.

Fourth, we consider $Y_{xd}=0$, $\tan\beta=20$, and $M_I=1.0\times 10^{11}$~GeV. We choose
three values for $M_V$: $M_V=400$~GeV, 1000~GeV,
and 2000~GeV. In Fig.~\ref{varying-with-M_S-yxd=0}, we present the upper bounds on
the lightest CP-even Higgs boson mass by varying $M_S$ from 360~GeV to 2~TeV.
As the value of $M_S$ increases, the upper bounds on the lightest 
CP-even Higgs boson mass increase from about 143~GeV to 162~GeV, 
from about 136~GeV to 150~GeV, and from about 134~GeV to 141~GeV, 
for $M_V=400$~GeV, 1000~GeV, and 2000~GeV, repectively.
Especially, to have the lightest 
CP-even Higgs boson mass upper bounds larger than 146 GeV,
we obtain that $M_S$ is larger than about $430$~GeV and 1260~GeV
for  $M_V=400$~GeV and 1000~GeV, respectively.
Moreover, the maximal Yukawa couplings $Y_{xu}$ 
decrease from about 1.049 to 1.007 for $M_S$ 
from 360~GeV to 2~TeV.

\begin{figure}[t]
      \begin{center}
            \includegraphics[width=6in]{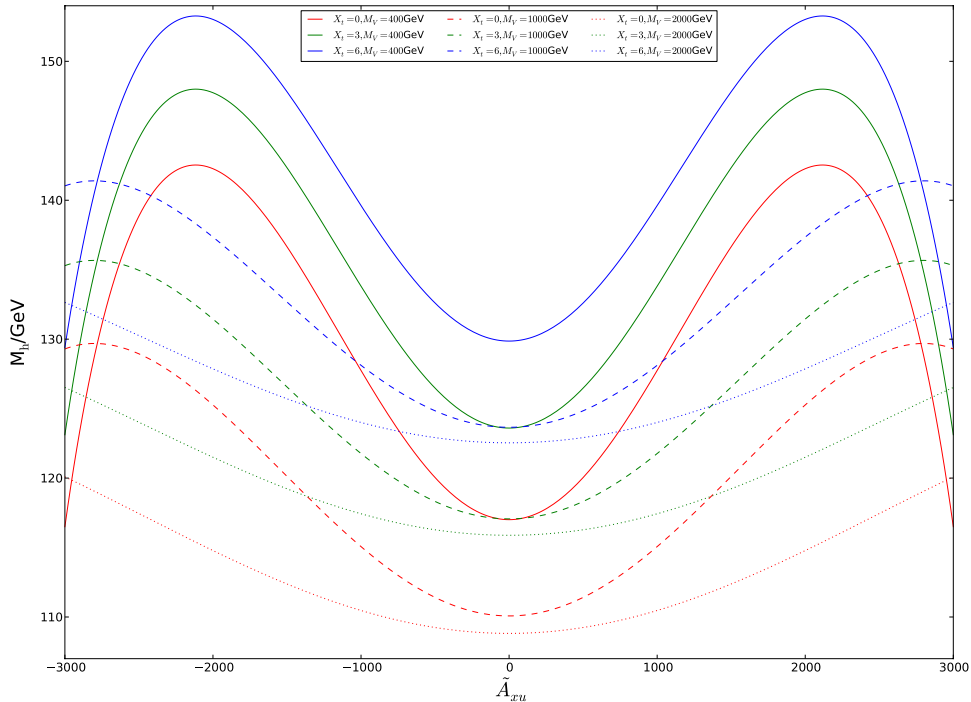}
      \end{center}
\vspace{-1cm}\caption{(color online). The upper bounds on the lightest
CP-even Higgs boson mass versus $X_{xu}$ in Model I with 
$Y_{xd}=0$, $\tan\beta=20$,  $M_S=800$~GeV, 
$M_V=400$~GeV, 1000~GeV, and 2000~GeV, and $X_t=0, ~3, ~{\rm and} ~6$.}
\label{varying-with-X_xu-yxd=0}
\end{figure}

Fifth, we consider $Y_{xd}=0$, $\tan\beta=20$, and $M_S=800$~GeV. Also, we choose
three values for $M_V$: $M_V=400$~GeV, 1000~GeV,
and 2000~GeV, and three values for $X_t$:
$X_t=0$, 3, and 6. For simplicity, we only consider
Model I here. In Fig.~\ref{varying-with-X_xu-yxd=0},
we present the upper bounds on the lightest CP-even
Higgs boson mass by varying ${\tilde A}_{xu}$. As we expected, they
behave just like the variations of the lightest CP-even Higgs boson
 mass upper bounds with varying stop
mixing $X_t$, which have been studied extensively before
in Refs.~\cite{Espinosa:1999zm,Espinosa:2000df,Carena:2000dp,Heinemeyer:2004ms}.












\section{Conclusion}

We calculated the lightest CP-even Higgs boson mass in
five kinds of testable flipped $SU(5)\times U(1)_X$
models from F-theory. Two kinds of models have 
vector-like particles around the TeV scale, while the 
other three kinds  also have the vector-like 
particles at the intermediate scale as the messenger
fields in gauge mediation. The Yukawa couplings 
for the TeV-scale vector-like particles and the 
third family of the SM fermions are required to be
smaller than three from the EW scale to 
the scale $M_{32}$. With the two-loop RGE running 
for both the gauge couplings and Yukawa couplings,
we obtained the maximal Yukawa couplings between
the TeV-scale vector-like particles and Higgs fields.
To calculate the lightest CP-even Higgs boson mass 
upper bounds, we used the RG improved effective Higgs 
potential approach, and considered the two-loop 
leading contributions in the MSSM and one-loop 
contributions from the TeV-scale vector-like particles. 
For simplicity, we assumed that the mixings both between 
the stops and between the TeV-scale vector-like scalars are 
maximal. The numerical results for these five kinds of 
models are roughly the same. With $M_V$ and $M_S$ around 
1~TeV,  we showed that the lightest CP-even Higgs boson 
mass can be close to 146 GeV naturally, which is the
upper bound from the current CMS and ATLAS
collaborations.



\begin{acknowledgments}

This research was supported in part 
by the Natural Science Foundation of China 
under grant numbers 10821504 and 11075194 (YH, TL, CT),
and by the DOE grant DE-FG03-95-Er-40917 (TL and DVN).

\end{acknowledgments}

\appendix



\section{Renormalization Group Equations in the SM with Vector-Like Particles}
When $M_V<M_S$, at the renormalization scale between them,  we have
the  Standard Model plus vector-like particles, with the RGE's
for the gauge couplings and Yukawa couplings as
follows~\cite{Cheng:1973nv,Vaughn:1981qi,Machacek:1983tz,Machacek:1983fi}:
\begin{eqnarray}
(4\pi)^2\frac{d}{dt}g_i=b_ig_i^3+\frac{g_i^3}{(4\pi)^2}\left[\sum_{j=1}^3B_{ij}g_j^2-\sum_{\alpha=u,d,e,xu,xd}\!\!\!\!\!\!\!\!d_i^{\alpha}\mbox{Tr}(Y_{\alpha}^{\dag}Y_{\alpha})\right],
\end{eqnarray}
where $t=\ln\mu$ and $\mu$ is the renormalization scale. The $g_1$,
$g_2$ and $g_3$ are the gauge couplings for $U(1)_Y$, $SU(2)_L$ and
$SU(3)_C$, respectively, where we use the $SU(5)$ normalization
$g_1^2\equiv(5/3)g_Y^2$. The beta-function coefficients are
\begin{eqnarray}
&&b=\left(
                                 \begin{array}{ccc}
                                   \frac{41}{10}, &  -\frac{19}{6}, & -7\\
                                 \end{array}
                               \right),~~~~~~~~~B=\left(
                                 \begin{array}{ccc}
                                   \frac{199}{50} & \frac{27}{10}  & \frac{44}{5}\\
                                    \frac{9}{10} & \frac{35}{6} & 12\\
                                    \frac{11}{10}& \frac{9}{2} & -26\\
                                 \end{array}
                               \right),\\
&&d^u=\left(\begin{array}{ccc}
        \frac{17}{10}, &  \frac{3}{2}, & 2\\
        \end{array}
        \right),~~~~d^d=d^{xu}=d^{xd}=\left(\begin{array}{ccc}
        \frac{1}{2}, &  \frac{3}{2}, & 2\\
        \end{array}
        \right),~~~~d^e=\left(\begin{array}{ccc}
        \frac{3}{2}, &  \frac{1}{2}, & 2\\
        \end{array}
        \right).
\end{eqnarray}
And
\begin{eqnarray}
\frac{d}{dt}Y_{u,d,e,xu,xd}=\frac{1}{16\pi^2}Y_{u,d,e,xu,xd}\beta^{(1)}_{u,d,e,xu,xd},
\end{eqnarray}
where
\begin{eqnarray}
&&\beta^{(1)}_{u}=\frac{3}{2}(Y_u^\dagger Y_u-Y_d^\dagger
Y_d)+Y_2-(\frac{17}{20}g^2_1+\frac{9}{4}g^2_2+8g_3^2),\\
&&\beta^{(1)}_{d}=\frac{3}{2}(Y_d^\dagger Y_d-Y_u^\dagger
Y_u)+Y_2-(\frac{1}{4}g^2_1+\frac{9}{4}g^2_2+8g_3^2),\\
&&\beta^{(1)}_{e}=\frac{3}{2}Y_e^\dagger Y_e+Y_2-\frac{9}{4}(g^2_1+g^2_2),\\
&&\beta^{(1)}_{xu}=\frac{3}{2}Y_{xu}^\dagger Y_{xu}+Y_2-(\frac{1}{4}g^2_1+\frac{9}{4}g^2_2+8g_3^2),\\
&&\beta^{(1)}_{xd}=\frac{3}{2}Y_{xd}^\dagger
Y_{xd}+Y_2-(\frac{1}{4}g^2_1+\frac{9}{4}g^2_2+8g_3^2),
\end{eqnarray}
with
\begin{eqnarray}
Y_2=Tr\{3Y_{u}^\dagger Y_{u}+3Y_{d}^\dagger Y_{d}+Y_{e}^\dagger
Y_{e}\}+3Y_{xu}^\dagger Y_{xu}+3Y_{xd}^\dagger Y_{xd}.
\end{eqnarray}





\section{Renormalization Group Equations in Model I}

In the Model I,
the two-loop renormalization group equations for the gauge couplings are
\begin{eqnarray}
(4\pi)^2\frac{d}{dt}~ g_i &=& b_i g_i^3
 +\frac{g_i^3}{(4\pi)^2}
\left[~ \sum_{j=1}^3 B_{ij}  g_j^2-\sum_{\alpha=u,d,e,xu, xd}d_i^\alpha
{\rm Tr}\left( Y_{\alpha}^{\dagger}Y_{\alpha}\right) \right] ~,~\,
\label{SUSYgauge}
\end{eqnarray}
where $Y_u$, $Y_d$, $Y_e$, $Y_{xu}$, and $Y_{xd}$  are the Yukawa couplings
for the up-type quark, down-type quark, lepton,  vector-like particles
$\overline{XF}$, and vector-like particles $XF$,
respectively. The beta-function coefficients are
\begin{eqnarray}
b=\left(\frac{33}{5},1,-3\right) + \left(\frac{3}{5}, 3, 3\right)~,~
\end{eqnarray}
\begin{eqnarray}
 B=\pmatrix{\frac{199}{25} &
\frac{27}{5}&\frac{88}{5}\cr \frac{9}{5} & 25&24 \cr
\frac{11}{5}&9&14} + \pmatrix{\frac{3}{25}&
\frac{3}{5}&\frac{16}{5}\cr \frac{1}{5} & 21 & 16 \cr
\frac{2}{5}& 6 &34}~,~
\end{eqnarray}
\begin{eqnarray}
&&d^u=\left(\frac{26}{5},6,4\right) ~,~
d^d=\left(\frac{14}{5},6,4\right) ~,~
d^e=\left(\frac{18}{5},2,0\right) ~,~ \\
&& d^{xu}=\left(\frac{14}{5},6,4\right) ~,~
d^{xd}=\left(\frac{14}{5},6,4\right) ~.~\,
\end{eqnarray}

The two-loop renormalization group equations for Yukawa couplings are
\begin{eqnarray}
(4\pi)^2\frac{d}{dt}~Y_{\alpha} &=& {1\over {16\pi^2}} \beta_{Y_\alpha}^{(1)}
+{1\over {(16\pi^2)^2}} \beta_{Y_\alpha}^{(2)} ~,~\,
\end{eqnarray}
where $\alpha= u, ~d, ~e, ~xu,~xd$. In addition, $\beta_{Y_\alpha}^{(1)}$
and $\beta_{Y_\alpha}^{(2)}$ are given as follows
\begin{eqnarray}
\beta_{Y_u}^{(1)} &=& Y_u \left(3{\rm Tr}(Y_u Y_u^{\dagger})+3Y_u^{\dagger} Y_u
+Y_d^{\dagger} Y_d + 3 Y_{xu}^{\dagger} Y_{xu} -{{16}\over 3}g_3^2-3 g_2^2
-{{13}\over {15}} g_1^2 \right)~,~\,
\end{eqnarray}
\begin{eqnarray}
\beta_{Y_u}^{(2)} &=& Y_u \left(-3{\rm Tr}(3Y_u Y_u^{\dagger}Y_u Y_u^{\dagger}
+Y_u Y_d^{\dagger}Y_d Y_u^{\dagger})
- 9 Y_{xu}^{\dagger}Y_{xu} Y_{xu}^{\dagger}Y_{xu}
- 9 Y_u^{\dagger}Y_u {\rm Tr}(Y_u Y_u^{\dagger})
\right. \nonumber \\ && \left.
- 9 Y_u^{\dagger}Y_u Y_{xu}^{\dagger}Y_{xu}
- Y_d^{\dagger}Y_d {\rm Tr}(3Y_d Y_d^{\dagger}+Y_e Y_e^{\dagger})
- 3 Y_d^{\dagger}Y_d Y_{xd} Y_{xd}^{\dagger}
\right. \nonumber \\ && \left.
-4 Y_u^{\dagger}Y_u Y_u^{\dagger}Y_u
-2 Y_d^{\dagger}Y_d Y_d^{\dagger}Y_d
-2 Y_d^{\dagger}Y_d Y_u^{\dagger}Y_u
+(16g_3^2+{4\over 5} g_1^2) {\rm Tr}(Y_u Y_u^{\dagger})
\right. \nonumber \\ && \left.
+ (16g_3^2-{2\over 5} g_1^2)  Y_{xu}^{\dagger} Y_{xu}
+(6g_2^2+{2\over 5}g_1^2) Y_u^{\dagger}Y_u
+ {2\over 5} g_1^2 Y_d^{\dagger}Y_d
+{{128}\over 9} g_3^4+ 8g_3^2g_2^2
\right. \nonumber \\ && \left.
+ {{136}\over {45}}g_3^2 g_1^2
+ {{33}\over 2}g_2^4+ g_2^2 g_1^2 + {{2977}\over {450}}g_1^4\right)
~,~\,
\end{eqnarray}
\begin{eqnarray}
\beta_{Y_d}^{(1)} &=& Y_d \left({\rm Tr}(3Y_d Y_d^{\dagger} +
Y_e Y_e^{\dagger}) + 3Y_d^{\dagger} Y_d
+Y_u^{\dagger} Y_u + 3 Y_{xd}^{\dagger} Y_{xd} -{{16}\over 3}g_3^2-3 g_2^2
-{{7}\over {15}} g_1^2 \right)~,~\,
\end{eqnarray}
\begin{eqnarray}
\beta_{Y_d}^{(2)} &=& Y_d \left(-3{\rm Tr}(3Y_d Y_d^{\dagger}Y_d Y_d^{\dagger}
+Y_d Y_u^{\dagger}Y_u Y_d^{\dagger}+ Y_e Y_e^{\dagger}Y_e Y_e^{\dagger} )
- 9 Y_{xd}^{\dagger}Y_{xd} Y_{xd}^{\dagger}Y_{xd}
- 3 Y_u^{\dagger}Y_u {\rm Tr}(Y_u Y_u^{\dagger})
\right. \nonumber \\ && \left.
- 3 Y_u^{\dagger}Y_u Y_{xu} Y_{xu}^{\dagger}
- 3 Y_d^{\dagger}Y_d {\rm Tr}(3Y_d Y_d^{\dagger}+ Y_e Y_e^{\dagger})
- 9 Y_d^{\dagger}Y_d Y_{xd}^{\dagger}Y_{xd}
-4 Y_d^{\dagger}Y_d Y_d^{\dagger}Y_d
\right. \nonumber \\ && \left.
-2 Y_u^{\dagger}Y_u Y_u^{\dagger}Y_u
-2 Y_u^{\dagger}Y_u Y_d^{\dagger}Y_d
+(16g_3^2-{2\over 5} g_1^2) {\rm Tr}(Y_d Y_d^{\dagger})
+{6\over 5} g_1^2 {\rm Tr}(Y_e Y_e^{\dagger})
\right. \nonumber \\ && \left.
+ (16g_3^2-{2\over 5} g_1^2)  Y_{xd}^{\dagger} Y_{xd}
+(6g_2^2+{4\over 5}g_1^2) Y_d^{\dagger}Y_d
+ {4\over 5} g_1^2 Y_u^{\dagger}Y_u
+{{128}\over 9} g_3^4+ 8g_3^2g_2^2
\right. \nonumber \\ && \left.
+ {{8}\over {9}}g_3^2 g_1^2
+ {{33}\over 2}g_2^4+ g_2^2 g_1^2 + {{1561}\over {450}}g_1^4\right)
~,~\,
\end{eqnarray}
\begin{eqnarray}
\beta_{Y_e}^{(1)} &=& Y_e \left({\rm Tr}(3Y_d Y_d^{\dagger} +
Y_e Y_e^{\dagger}) + 3Y_e^{\dagger} Y_e
+ 3 Y_{xd}^{\dagger} Y_{xd} -3 g_2^2
-{{9}\over {5}} g_1^2 \right)~,~\,
\end{eqnarray}
\begin{eqnarray}
\beta_{Y_e}^{(2)} &=& Y_e \left(-3{\rm Tr}(3Y_d Y_d^{\dagger}Y_d Y_d^{\dagger}
+Y_d Y_u^{\dagger}Y_u Y_d^{\dagger}+ Y_e Y_e^{\dagger}Y_e Y_e^{\dagger} )
- 9 Y_{xd}^{\dagger}Y_{xd} Y_{xd}^{\dagger}Y_{xd}
\right. \nonumber \\ && \left.
- 3 Y_e^{\dagger}Y_e {\rm Tr}(3Y_d Y_d^{\dagger}+ Y_e Y_e^{\dagger})
- 9 Y_e^{\dagger}Y_e Y_{xd}^{\dagger}Y_{xd}
-4 Y_e^{\dagger}Y_e Y_e^{\dagger}Y_e
\right. \nonumber \\ && \left.
+(16g_3^2-{2\over 5} g_1^2) {\rm Tr}(Y_d Y_d^{\dagger})
+{6\over 5} g_1^2 {\rm Tr}(Y_e Y_e^{\dagger})
+ (16g_3^2-{2\over 5} g_1^2)  Y_{xd}^{\dagger} Y_{xd}
\right. \nonumber \\ && \left.
+6g_2^2 Y_e^{\dagger}Y_e
+ {{33}\over 2}g_2^4+ {9\over 5} g_2^2 g_1^2 + {{729}\over {50}}g_1^4\right)
~,~\,
\end{eqnarray}
\begin{eqnarray}
\beta_{Y_{xu}}^{(1)} &=& Y_{xu} \left(3{\rm Tr}(Y_u Y_u^{\dagger})
+ 6 Y_{xu}^{\dagger} Y_{xu} -{{16}\over 3}g_3^2-3 g_2^2
-{{7}\over {15}} g_1^2 \right)~,~\,
\end{eqnarray}
\begin{eqnarray}
\beta_{Y_{xu}}^{(2)} &=& Y_{xu} \left(-3{\rm Tr}(3Y_u Y_u^{\dagger}Y_u Y_u^{\dagger}
+Y_u Y_d^{\dagger}Y_d Y_u^{\dagger})
- 22 Y_{xu}^{\dagger}Y_{xu} Y_{xu}^{\dagger}Y_{xu}
- 9 Y_{xu}^{\dagger}Y_{xu} {\rm Tr}(Y_u Y_u^{\dagger})
\right. \nonumber \\ && \left.
+(16g_3^2+{4\over 5} g_1^2) {\rm Tr}(Y_u Y_u^{\dagger})
+ (16g_3^2 +6g_2^2 +{2\over 5} g_1^2)  Y_{xu}^{\dagger} Y_{xu}
\right. \nonumber \\ && \left.
+{{128}\over 9} g_3^4+ 8g_3^2g_2^2
+ {{8}\over {9}}g_3^2 g_1^2
+ {{33}\over 2}g_2^4+ g_2^2 g_1^2 + {{1561}\over {450}}g_1^4\right)
~,~\,
\end{eqnarray}
\begin{eqnarray}
\beta_{Y_{xd}}^{(1)} &=& Y_{xd} \left({\rm Tr}(3Y_{d} Y_{d}^{\dagger} +
Y_e Y_e^{\dagger}) + 6 Y_{xd}^{\dagger} Y_{xd} -{{16}\over 3}g_3^2-3 g_2^2
-{{7}\over {15}} g_1^2 \right)~,~\,
\end{eqnarray}
\begin{eqnarray}
\beta_{Y_{xd}}^{(2)} &=& Y_{xd} \left(-3{\rm Tr}(3Y_d Y_d^{\dagger}Y_d Y_d^{\dagger}
+Y_d Y_u^{\dagger}Y_u Y_d^{\dagger}+ Y_e Y_e^{\dagger}Y_e Y_e^{\dagger} )
- 22 Y_{xd}^{\dagger}Y_{xd} Y_{xd}^{\dagger}Y_{xd}
\right. \nonumber \\ && \left.
- 3 Y_{xd}^{\dagger}Y_{xd} {\rm Tr}(3Y_d Y_d^{\dagger}+ Y_e Y_e^{\dagger})
+(16g_3^2-{2\over 5} g_1^2) {\rm Tr}(Y_d Y_d^{\dagger})
+{6\over 5} g_1^2 {\rm Tr}(Y_e Y_e^{\dagger})
\right. \nonumber \\ && \left.
+ (16g_3^2+ 6g_2^2+{2\over 5} g_1^2)  Y_{xd}^{\dagger} Y_{xd}
+{{128}\over 9} g_3^4+ 8g_3^2g_2^2
+ {{8}\over {9}}g_3^2 g_1^2
\right. \nonumber \\ && \left.
+ {{33}\over 2}g_2^4+ g_2^2 g_1^2 + {{1561}\over {450}}g_1^4\right)
~.~\,
\end{eqnarray}



\section{Renormalization Group Equations in Model II}

In Model II, below the intermediate scale $M_{I}=1.0\times 10^{11}$ GeV, we have the
same RGEs as in Model I. Above $M_{I}$, we have additional vector-like particles
($Xf$, $\overline{Xf}$). Thus, we need to add extra contributions to
$b$ and $B$ from the vector-like particles
($Xf$, $\overline{Xf}$). Comparing to the RGEs in Model I,
we also need to change the coefficients of the $g_3^4$, $g_2^4$ and
$g_1^4$ terms in $\beta_{Y_u}^{(2)}$, $\beta_{Y_d}^{(2)}$,
$\beta_{Y_{xu}}^{(2)}$, and $\beta_{Y_{xd}}^{(2)}$, and
change the coefficients of the $g_2^4$ and
$g_1^4$ terms in $\beta_{Y_e}^{(2)}$.
In short, comparing to the RGEs in Model I,
the coefficients in the RGEs above $M_{I}$, which need to be changed,
are the following:
\begin{eqnarray}
b=\left(\frac{33}{5},1,-3\right) + \left(\frac{3}{5}, 3, 3\right)
+ \left(\frac{11}{5}, 1, 1\right)
~,~
\end{eqnarray}
\begin{eqnarray}
 B=\pmatrix{\frac{199}{25} &
\frac{27}{5}&\frac{88}{5}\cr \frac{9}{5} & 25&24 \cr
\frac{11}{5}&9&14} + \pmatrix{\frac{3}{25}&
\frac{3}{5}&\frac{16}{5}\cr \frac{1}{5} & 21 & 16 \cr
\frac{2}{5}& 6 &34}
+\pmatrix{\frac{31}{15}& \frac{9}{5}
 &\frac{128}{15}\cr \frac{3}{5} & 7 & 0 \cr
\frac{16}{15}& 0 & \frac{34}{3}}~,~\,
\end{eqnarray}
\begin{eqnarray}
\beta_{Y_u}^{(2)} &=& Y_u \left(-3{\rm Tr}(3Y_u Y_u^{\dagger}Y_u Y_u^{\dagger}
+Y_u Y_d^{\dagger}Y_d Y_u^{\dagger})
- 9 Y_{xu}^{\dagger}Y_{xu} Y_{xu}^{\dagger}Y_{xu}
- 9 Y_u^{\dagger}Y_u {\rm Tr}(Y_u Y_u^{\dagger})
\right. \nonumber \\ && \left.
- 9 Y_u^{\dagger}Y_u Y_{xu}^{\dagger}Y_{xu}
- Y_d^{\dagger}Y_d {\rm Tr}(3Y_d Y_d^{\dagger}+Y_e Y_e^{\dagger})
- 3 Y_d^{\dagger}Y_d Y_{xd} Y_{xd}^{\dagger}
\right. \nonumber \\ && \left.
-4 Y_u^{\dagger}Y_u Y_u^{\dagger}Y_u
-2 Y_d^{\dagger}Y_d Y_d^{\dagger}Y_d
-2 Y_d^{\dagger}Y_d Y_u^{\dagger}Y_u
+(16g_3^2+{4\over 5} g_1^2) {\rm Tr}(Y_u Y_u^{\dagger})
\right. \nonumber \\ && \left.
+ (16g_3^2-{2\over 5} g_1^2)  Y_{xu}^{\dagger} Y_{xu}
+(6g_2^2+{2\over 5}g_1^2) Y_u^{\dagger}Y_u
+ {2\over 5} g_1^2 Y_d^{\dagger}Y_d
+{{176}\over 9} g_3^4+ 8g_3^2g_2^2
\right. \nonumber \\ && \left.
+ {{136}\over {45}}g_3^2 g_1^2
+ {{39}\over 2}g_2^4+ g_2^2 g_1^2 + {{767}\over {90}}g_1^4\right)
~,~\,
\end{eqnarray}
\begin{eqnarray}
\beta_{Y_d}^{(2)} &=& Y_d \left(-3{\rm Tr}(3Y_d Y_d^{\dagger}Y_d Y_d^{\dagger}
+Y_d Y_u^{\dagger}Y_u Y_d^{\dagger}+ Y_e Y_e^{\dagger}Y_e Y_e^{\dagger} )
- 9 Y_{xd}^{\dagger}Y_{xd} Y_{xd}^{\dagger}Y_{xd}
- 3 Y_u^{\dagger}Y_u {\rm Tr}(Y_u Y_u^{\dagger})
\right. \nonumber \\ && \left.
- 3 Y_u^{\dagger}Y_u Y_{xu} Y_{xu}^{\dagger}
- 3 Y_d^{\dagger}Y_d {\rm Tr}(3Y_d Y_d^{\dagger}+ Y_e Y_e^{\dagger})
- 9 Y_d^{\dagger}Y_d Y_{xd}^{\dagger}Y_{xd}
-4 Y_d^{\dagger}Y_d Y_d^{\dagger}Y_d
\right. \nonumber \\ && \left.
-2 Y_u^{\dagger}Y_u Y_u^{\dagger}Y_u
-2 Y_u^{\dagger}Y_u Y_d^{\dagger}Y_d
+(16g_3^2-{2\over 5} g_1^2) {\rm Tr}(Y_d Y_d^{\dagger})
+{6\over 5} g_1^2 {\rm Tr}(Y_e Y_e^{\dagger})
\right. \nonumber \\ && \left.
+ (16g_3^2-{2\over 5} g_1^2)  Y_{xd}^{\dagger} Y_{xd}
+(6g_2^2+{4\over 5}g_1^2) Y_d^{\dagger}Y_d
+ {4\over 5} g_1^2 Y_u^{\dagger}Y_u
+{{176}\over 9} g_3^4+ 8g_3^2g_2^2
\right. \nonumber \\ && \left.
+ {{8}\over {9}}g_3^2 g_1^2
+ {{39}\over 2}g_2^4+ g_2^2 g_1^2 + {{2023}\over {450}}g_1^4\right)
~,~\,
\end{eqnarray}
\begin{eqnarray}
\beta_{Y_e}^{(2)} &=& Y_e \left(-3{\rm Tr}(3Y_d Y_d^{\dagger}Y_d Y_d^{\dagger}
+Y_d Y_u^{\dagger}Y_u Y_d^{\dagger}+ Y_e Y_e^{\dagger}Y_e Y_e^{\dagger} )
- 9 Y_{xd}^{\dagger}Y_{xd} Y_{xd}^{\dagger}Y_{xd}
\right. \nonumber \\ && \left.
- 3 Y_e^{\dagger}Y_e {\rm Tr}(3Y_d Y_d^{\dagger}+ Y_e Y_e^{\dagger})
- 9 Y_e^{\dagger}Y_e Y_{xd}^{\dagger}Y_{xd}
-4 Y_e^{\dagger}Y_e Y_e^{\dagger}Y_e
\right. \nonumber \\ && \left.
+(16g_3^2-{2\over 5} g_1^2) {\rm Tr}(Y_d Y_d^{\dagger})
+{6\over 5} g_1^2 {\rm Tr}(Y_e Y_e^{\dagger})
+ (16g_3^2-{2\over 5} g_1^2)  Y_{xd}^{\dagger} Y_{xd}
\right. \nonumber \\ && \left.
+6g_2^2 Y_e^{\dagger}Y_e
+ {{39}\over 2}g_2^4+ {9\over 5} g_2^2 g_1^2 + {{927}\over {50}}g_1^4\right)
~,~\,
\end{eqnarray}
\begin{eqnarray}
\beta_{Y_{xu}}^{(2)} &=& Y_{xu} \left(-3{\rm Tr}(3Y_u Y_u^{\dagger}Y_u Y_u^{\dagger}
+Y_u Y_d^{\dagger}Y_d Y_u^{\dagger})
- 22 Y_{xu}^{\dagger}Y_{xu} Y_{xu}^{\dagger}Y_{xu}
- 9 Y_{xu}^{\dagger}Y_{xu} {\rm Tr}(Y_u Y_u^{\dagger})
\right. \nonumber \\ && \left.
+(16g_3^2+{4\over 5} g_1^2) {\rm Tr}(Y_u Y_u^{\dagger})
+ (16g_3^2 +6g_2^2 +{2\over 5} g_1^2)  Y_{xu}^{\dagger} Y_{xu}
\right. \nonumber \\ && \left.
+{{176}\over 9} g_3^4+ 8g_3^2g_2^2
+ {{8}\over {9}}g_3^2 g_1^2
+ {{39}\over 2}g_2^4+ g_2^2 g_1^2 + {{2023}\over {450}}g_1^4\right)
~,~\,
\end{eqnarray}
\begin{eqnarray}
\beta_{Y_{xd}}^{(2)} &=& Y_{xd} \left(-3{\rm Tr}(3Y_d Y_d^{\dagger}Y_d Y_d^{\dagger}
+Y_d Y_u^{\dagger}Y_u Y_d^{\dagger}+ Y_e Y_e^{\dagger}Y_e Y_e^{\dagger} )
- 22 Y_{xd}^{\dagger}Y_{xd} Y_{xd}^{\dagger}Y_{xd}
\right. \nonumber \\ && \left.
- 3 Y_{xd}^{\dagger}Y_{xd} {\rm Tr}(3Y_d Y_d^{\dagger}+ Y_e Y_e^{\dagger})
+(16g_3^2-{2\over 5} g_1^2) {\rm Tr}(Y_d Y_d^{\dagger})
+{6\over 5} g_1^2 {\rm Tr}(Y_e Y_e^{\dagger})
\right. \nonumber \\ && \left.
+ (16g_3^2+ 6g_2^2+{2\over 5} g_1^2)  Y_{xd}^{\dagger} Y_{xd}
+{{176}\over 9} g_3^4+ 8g_3^2g_2^2
+ {{8}\over {9}}g_3^2 g_1^2
\right. \nonumber \\ && \left.
+ {{39}\over 2}g_2^4+ g_2^2 g_1^2 + {{2023}\over {450}}g_1^4\right)
~.~\,
\end{eqnarray}





\section{Renormalization Group Equations in Model III}

In Model III, below the intermediate scale $M_{I}=1.0\times 10^{11}$ GeV, we have the
same RGEs as in Model I. Above $M_{I}$, we have additional vector-like particles
($Xf$, $\overline{Xf}$) and ($Xl$, $\overline{Xl}$). 
Thus, we need to add extra contributions to
$b$ and $B$ from the vector-like particles
($Xf$, $\overline{Xf}$) and ($Xl$, $\overline{Xl}$). 
Comparing to the RGEs in Model I,
we also need to change the coefficients of the $g_3^4$, $g_2^4$ and
$g_1^4$ terms in $\beta_{Y_u}^{(2)}$, $\beta_{Y_d}^{(2)}$,
$\beta_{Y_{xu}}^{(2)}$, and $\beta_{Y_{xd}}^{(2)}$, and
change the coefficients of the $g_2^4$ and
$g_1^4$ terms in $\beta_{Y_e}^{(2)}$.
In short, comparing to the RGEs in Model I,
the coefficients in the RGEs above $M_{I}$, which need to be changed,
are the following:
\begin{eqnarray}
b=\left(\frac{33}{5},1,-3\right) + \left(\frac{3}{5}, 3, 3\right)
+ \left(\frac{17}{5}, 1, 1\right)
~,~
\end{eqnarray}
\begin{eqnarray}
 B=\pmatrix{\frac{199}{25} &
\frac{27}{5}&\frac{88}{5}\cr \frac{9}{5} & 25&24 \cr
\frac{11}{5}&9&14} + \pmatrix{\frac{3}{25}&
\frac{3}{5}&\frac{16}{5}\cr \frac{1}{5} & 21 & 16 \cr
\frac{2}{5}& 6 &34}
+\pmatrix{\frac{371}{15}& \frac{9}{5}
 &\frac{128}{15}\cr \frac{3}{5} & 7 & 0 \cr
\frac{16}{15}& 0 & \frac{34}{3}}~,~\,
\end{eqnarray}
\begin{eqnarray}
\beta_{Y_u}^{(2)} &=& Y_u \left(-3{\rm Tr}(3Y_u Y_u^{\dagger}Y_u Y_u^{\dagger}
+Y_u Y_d^{\dagger}Y_d Y_u^{\dagger})
- 9 Y_{xu}^{\dagger}Y_{xu} Y_{xu}^{\dagger}Y_{xu}
- 9 Y_u^{\dagger}Y_u {\rm Tr}(Y_u Y_u^{\dagger})
\right. \nonumber \\ && \left.
- 9 Y_u^{\dagger}Y_u Y_{xu}^{\dagger}Y_{xu}
- Y_d^{\dagger}Y_d {\rm Tr}(3Y_d Y_d^{\dagger}+Y_e Y_e^{\dagger})
- 3 Y_d^{\dagger}Y_d Y_{xd} Y_{xd}^{\dagger}
\right. \nonumber \\ && \left.
-4 Y_u^{\dagger}Y_u Y_u^{\dagger}Y_u
-2 Y_d^{\dagger}Y_d Y_d^{\dagger}Y_d
-2 Y_d^{\dagger}Y_d Y_u^{\dagger}Y_u
+(16g_3^2+{4\over 5} g_1^2) {\rm Tr}(Y_u Y_u^{\dagger})
\right. \nonumber \\ && \left.
+ (16g_3^2-{2\over 5} g_1^2)  Y_{xu}^{\dagger} Y_{xu}
+(6g_2^2+{2\over 5}g_1^2) Y_u^{\dagger}Y_u
+ {2\over 5} g_1^2 Y_d^{\dagger}Y_d
+{{176}\over 9} g_3^4+ 8g_3^2g_2^2
\right. \nonumber \\ && \left.
+ {{136}\over {45}}g_3^2 g_1^2
+ {{39}\over 2}g_2^4+ g_2^2 g_1^2 + {{4303}\over {450}}g_1^4\right)
~,~\,
\end{eqnarray}
\begin{eqnarray}
\beta_{Y_d}^{(2)} &=& Y_d \left(-3{\rm Tr}(3Y_d Y_d^{\dagger}Y_d Y_d^{\dagger}
+Y_d Y_u^{\dagger}Y_u Y_d^{\dagger}+ Y_e Y_e^{\dagger}Y_e Y_e^{\dagger} )
- 9 Y_{xd}^{\dagger}Y_{xd} Y_{xd}^{\dagger}Y_{xd}
- 3 Y_u^{\dagger}Y_u {\rm Tr}(Y_u Y_u^{\dagger})
\right. \nonumber \\ && \left.
- 3 Y_u^{\dagger}Y_u Y_{xu} Y_{xu}^{\dagger}
- 3 Y_d^{\dagger}Y_d {\rm Tr}(3Y_d Y_d^{\dagger}+ Y_e Y_e^{\dagger})
- 9 Y_d^{\dagger}Y_d Y_{xd}^{\dagger}Y_{xd}
-4 Y_d^{\dagger}Y_d Y_d^{\dagger}Y_d
\right. \nonumber \\ && \left.
-2 Y_u^{\dagger}Y_u Y_u^{\dagger}Y_u
-2 Y_u^{\dagger}Y_u Y_d^{\dagger}Y_d
+(16g_3^2-{2\over 5} g_1^2) {\rm Tr}(Y_d Y_d^{\dagger})
+{6\over 5} g_1^2 {\rm Tr}(Y_e Y_e^{\dagger})
\right. \nonumber \\ && \left.
+ (16g_3^2-{2\over 5} g_1^2)  Y_{xd}^{\dagger} Y_{xd}
+(6g_2^2+{4\over 5}g_1^2) Y_d^{\dagger}Y_d
+ {4\over 5} g_1^2 Y_u^{\dagger}Y_u
+{{176}\over 9} g_3^4+ 8g_3^2g_2^2
\right. \nonumber \\ && \left.
+ {{8}\over {9}}g_3^2 g_1^2
+ {{39}\over 2}g_2^4+ g_2^2 g_1^2 + {{91}\over {18}}g_1^4\right)
~,~\,
\end{eqnarray}
\begin{eqnarray}
\beta_{Y_e}^{(2)} &=& Y_e \left(-3{\rm Tr}(3Y_d Y_d^{\dagger}Y_d Y_d^{\dagger}
+Y_d Y_u^{\dagger}Y_u Y_d^{\dagger}+ Y_e Y_e^{\dagger}Y_e Y_e^{\dagger} )
- 9 Y_{xd}^{\dagger}Y_{xd} Y_{xd}^{\dagger}Y_{xd}
\right. \nonumber \\ && \left.
- 3 Y_e^{\dagger}Y_e {\rm Tr}(3Y_d Y_d^{\dagger}+ Y_e Y_e^{\dagger})
- 9 Y_e^{\dagger}Y_e Y_{xd}^{\dagger}Y_{xd}
-4 Y_e^{\dagger}Y_e Y_e^{\dagger}Y_e
\right. \nonumber \\ && \left.
+(16g_3^2-{2\over 5} g_1^2) {\rm Tr}(Y_d Y_d^{\dagger})
+{6\over 5} g_1^2 {\rm Tr}(Y_e Y_e^{\dagger})
+ (16g_3^2-{2\over 5} g_1^2)  Y_{xd}^{\dagger} Y_{xd}
\right. \nonumber \\ && \left.
+6g_2^2 Y_e^{\dagger}Y_e
+ {{39}\over 2}g_2^4+ {9\over 5} g_2^2 g_1^2 + {{207}\over {10}}g_1^4\right)
~,~\,
\end{eqnarray}
\begin{eqnarray}
\beta_{Y_{xu}}^{(2)} &=& Y_{xu} \left(-3{\rm Tr}(3Y_u Y_u^{\dagger}Y_u Y_u^{\dagger}
+Y_u Y_d^{\dagger}Y_d Y_u^{\dagger})
- 22 Y_{xu}^{\dagger}Y_{xu} Y_{xu}^{\dagger}Y_{xu}
- 9 Y_{xu}^{\dagger}Y_{xu} {\rm Tr}(Y_u Y_u^{\dagger})
\right. \nonumber \\ && \left.
+(16g_3^2+{4\over 5} g_1^2) {\rm Tr}(Y_u Y_u^{\dagger})
+ (16g_3^2 +6g_2^2 +{2\over 5} g_1^2)  Y_{xu}^{\dagger} Y_{xu}
\right. \nonumber \\ && \left.
+{{176}\over 9} g_3^4+ 8g_3^2g_2^2
+ {{8}\over {9}}g_3^2 g_1^2
+ {{39}\over 2}g_2^4+ g_2^2 g_1^2 + {{91}\over {18}}g_1^4\right)
~,~\,
\end{eqnarray}
\begin{eqnarray}
\beta_{Y_{xd}}^{(2)} &=& Y_{xd} \left(-3{\rm Tr}(3Y_d Y_d^{\dagger}Y_d Y_d^{\dagger}
+Y_d Y_u^{\dagger}Y_u Y_d^{\dagger}+ Y_e Y_e^{\dagger}Y_e Y_e^{\dagger} )
- 22 Y_{xd}^{\dagger}Y_{xd} Y_{xd}^{\dagger}Y_{xd}
\right. \nonumber \\ && \left.
- 3 Y_{xd}^{\dagger}Y_{xd} {\rm Tr}(3Y_d Y_d^{\dagger}+ Y_e Y_e^{\dagger})
+(16g_3^2-{2\over 5} g_1^2) {\rm Tr}(Y_d Y_d^{\dagger})
+{6\over 5} g_1^2 {\rm Tr}(Y_e Y_e^{\dagger})
\right. \nonumber \\ && \left.
+ (16g_3^2+ 6g_2^2+{2\over 5} g_1^2)  Y_{xd}^{\dagger} Y_{xd}
+{{176}\over 9} g_3^4+ 8g_3^2g_2^2
+ {{8}\over {9}}g_3^2 g_1^2
\right. \nonumber \\ && \left.
+ {{39}\over 2}g_2^4+ g_2^2 g_1^2 + {{91}\over {18}}g_1^4\right)
~.~\,
\end{eqnarray}









\section{Renormalization Group Equations in Model IV}

In Model IV,  we have additional vector-like particles
$Xl$ and $\overline{Xl}$. Thus, we need to add extra contributions to
$b$ and $B$ from the vector-like particles
$Xl$ and $\overline{Xl}$. Comparing to the RGEs in Model I,
we also need to change the coefficients of the
$g_1^4$ terms in $\beta_{Y_u}^{(2)}$, $\beta_{Y_d}^{(2)}$,
$\beta_{Y_e}^{(2)}$, $\beta_{Y_{xu}}^{(2)}$,  and  $\beta_{Y_{xd}}^{(2)}$.
In short, comparing to the RGEs in
 Model I,  the corresponding coefficients of the
RGEs, which need to be changed,
are the following:
\begin{eqnarray}
b=\left(\frac{33}{5},1,-3\right) + \left(\frac{3}{5}, 3, 3\right)
+ \left(\frac{6}{5}, 0, 0\right)
~,~
\end{eqnarray}
\begin{eqnarray}
 B=\pmatrix{\frac{199}{25} &
\frac{27}{5}&\frac{88}{5}\cr \frac{9}{5} & 25&24 \cr
\frac{11}{5}&9&14} + \pmatrix{\frac{3}{25}&
\frac{3}{5}&\frac{16}{5}\cr \frac{1}{5} & 21 & 16 \cr
\frac{2}{5}& 6 &34}
+\pmatrix{\frac{36}{25}&
0 & 0 \cr 0 & 0 & 0 \cr
0 & 0 & 0 }~,~\,
\end{eqnarray}
\begin{eqnarray}
\beta_{Y_u}^{(2)} &=& Y_u \left(-3{\rm Tr}(3Y_u Y_u^{\dagger}Y_u Y_u^{\dagger}
+Y_u Y_d^{\dagger}Y_d Y_u^{\dagger})
- 9 Y_{xu}^{\dagger}Y_{xu} Y_{xu}^{\dagger}Y_{xu}
- 9 Y_u^{\dagger}Y_u {\rm Tr}(Y_u Y_u^{\dagger})
\right. \nonumber \\ && \left.
- 9 Y_u^{\dagger}Y_u Y_{xu}^{\dagger}Y_{xu}
- Y_d^{\dagger}Y_d {\rm Tr}(3Y_d Y_d^{\dagger}+Y_e Y_e^{\dagger})
- 3 Y_d^{\dagger}Y_d Y_{xd} Y_{xd}^{\dagger}
\right. \nonumber \\ && \left.
-4 Y_u^{\dagger}Y_u Y_u^{\dagger}Y_u
-2 Y_d^{\dagger}Y_d Y_d^{\dagger}Y_d
-2 Y_d^{\dagger}Y_d Y_u^{\dagger}Y_u
+(16g_3^2+{4\over 5} g_1^2) {\rm Tr}(Y_u Y_u^{\dagger})
\right. \nonumber \\ && \left.
+ (16g_3^2-{2\over 5} g_1^2)  Y_{xu}^{\dagger} Y_{xu}
+(6g_2^2+{2\over 5}g_1^2) Y_u^{\dagger}Y_u
+ {2\over 5} g_1^2 Y_d^{\dagger}Y_d
+{{128}\over 9} g_3^4+ 8g_3^2g_2^2
\right. \nonumber \\ && \left.
+ {{136}\over {45}}g_3^2 g_1^2
+ {{33}\over 2}g_2^4+ g_2^2 g_1^2 + {{689}\over {90}}g_1^4\right)
~,~\,
\end{eqnarray}
\begin{eqnarray}
\beta_{Y_d}^{(2)} &=& Y_d \left(-3{\rm Tr}(3Y_d Y_d^{\dagger}Y_d Y_d^{\dagger}
+Y_d Y_u^{\dagger}Y_u Y_d^{\dagger}+ Y_e Y_e^{\dagger}Y_e Y_e^{\dagger} )
- 9 Y_{xd}^{\dagger}Y_{xd} Y_{xd}^{\dagger}Y_{xd}
- 3 Y_u^{\dagger}Y_u {\rm Tr}(Y_u Y_u^{\dagger})
\right. \nonumber \\ && \left.
- 3 Y_u^{\dagger}Y_u Y_{xu} Y_{xu}^{\dagger}
- 3 Y_d^{\dagger}Y_d {\rm Tr}(3Y_d Y_d^{\dagger}+ Y_e Y_e^{\dagger})
- 9 Y_d^{\dagger}Y_d Y_{xd}^{\dagger}Y_{xd}
-4 Y_d^{\dagger}Y_d Y_d^{\dagger}Y_d
\right. \nonumber \\ && \left.
-2 Y_u^{\dagger}Y_u Y_u^{\dagger}Y_u
-2 Y_u^{\dagger}Y_u Y_d^{\dagger}Y_d
+(16g_3^2-{2\over 5} g_1^2) {\rm Tr}(Y_d Y_d^{\dagger})
+{6\over 5} g_1^2 {\rm Tr}(Y_e Y_e^{\dagger})
\right. \nonumber \\ && \left.
+ (16g_3^2-{2\over 5} g_1^2)  Y_{xd}^{\dagger} Y_{xd}
+(6g_2^2+{4\over 5}g_1^2) Y_d^{\dagger}Y_d
+ {4\over 5} g_1^2 Y_u^{\dagger}Y_u
+{{128}\over 9} g_3^4+ 8g_3^2g_2^2
\right. \nonumber \\ && \left.
+ {{8}\over {9}}g_3^2 g_1^2
+ {{33}\over 2}g_2^4+ g_2^2 g_1^2 + {{1813}\over {450}}g_1^4\right)
~,~\,
\end{eqnarray}
\begin{eqnarray}
\beta_{Y_e}^{(2)} &=& Y_e \left(-3{\rm Tr}(3Y_d Y_d^{\dagger}Y_d Y_d^{\dagger}
+Y_d Y_u^{\dagger}Y_u Y_d^{\dagger}+ Y_e Y_e^{\dagger}Y_e Y_e^{\dagger} )
- 9 Y_{xd}^{\dagger}Y_{xd} Y_{xd}^{\dagger}Y_{xd}
\right. \nonumber \\ && \left.
- 3 Y_e^{\dagger}Y_e {\rm Tr}(3Y_d Y_d^{\dagger}+ Y_e Y_e^{\dagger})
- 9 Y_e^{\dagger}Y_e Y_{xd}^{\dagger}Y_{xd}
-4 Y_e^{\dagger}Y_e Y_e^{\dagger}Y_e
\right. \nonumber \\ && \left.
+(16g_3^2-{2\over 5} g_1^2) {\rm Tr}(Y_d Y_d^{\dagger})
+{6\over 5} g_1^2 {\rm Tr}(Y_e Y_e^{\dagger})
+ (16g_3^2-{2\over 5} g_1^2)  Y_{xd}^{\dagger} Y_{xd}
\right. \nonumber \\ && \left.
+6g_2^2 Y_e^{\dagger}Y_e
+ {{33}\over 2}g_2^4+ {9\over 5} g_2^2 g_1^2 + {{837}\over {50}}g_1^4\right)
~,~\,
\end{eqnarray}
\begin{eqnarray}
\beta_{Y_{xu}}^{(2)} &=& Y_{xu} \left(-3{\rm Tr}(3Y_u Y_u^{\dagger}Y_u Y_u^{\dagger}
+Y_u Y_d^{\dagger}Y_d Y_u^{\dagger})
- 22 Y_{xu}^{\dagger}Y_{xu} Y_{xu}^{\dagger}Y_{xu}
- 9 Y_{xu}^{\dagger}Y_{xu} {\rm Tr}(Y_u Y_u^{\dagger})
\right. \nonumber \\ && \left.
+(16g_3^2+{4\over 5} g_1^2) {\rm Tr}(Y_u Y_u^{\dagger})
+ (16g_3^2 +6g_2^2 +{2\over 5} g_1^2)  Y_{xu}^{\dagger} Y_{xu}
\right. \nonumber \\ && \left.
+{{128}\over 9} g_3^4+ 8g_3^2g_2^2
+ {{8}\over {9}}g_3^2 g_1^2
+ {{33}\over 2}g_2^4+ g_2^2 g_1^2 + {{1813}\over {450}}g_1^4\right)
~,~\,
\end{eqnarray}
\begin{eqnarray}
\beta_{Y_{xd}}^{(2)} &=& Y_{xd} \left(-3{\rm Tr}(3Y_d Y_d^{\dagger}Y_d Y_d^{\dagger}
+Y_d Y_u^{\dagger}Y_u Y_d^{\dagger}+ Y_e Y_e^{\dagger}Y_e Y_e^{\dagger} )
- 22 Y_{xd}^{\dagger}Y_{xd} Y_{xd}^{\dagger}Y_{xd}
\right. \nonumber \\ && \left.
- 3 Y_{xd}^{\dagger}Y_{xd} {\rm Tr}(3Y_d Y_d^{\dagger}+ Y_e Y_e^{\dagger})
+(16g_3^2-{2\over 5} g_1^2) {\rm Tr}(Y_d Y_d^{\dagger})
+{6\over 5} g_1^2 {\rm Tr}(Y_e Y_e^{\dagger})
\right. \nonumber \\ && \left.
+ (16g_3^2+ 6g_2^2+{2\over 5} g_1^2)  Y_{xd}^{\dagger} Y_{xd}
+{{128}\over 9} g_3^4+ 8g_3^2g_2^2
+ {{8}\over {9}}g_3^2 g_1^2
\right. \nonumber \\ && \left.
+ {{33}\over 2}g_2^4+ g_2^2 g_1^2 + {{1813}\over {450}}g_1^4\right)
~.~\,
\end{eqnarray}







\section{Renormalization Group Equations in Model V}

In Model V, below the intermediate scale $M_{I}=1.0\times 10^{11}$ GeV, we have the
same RGEs as in Model IV. Above $M_{I}$, we have extra vector-like particles
($Xf$, $\overline{Xf}$), and  we have the
same RGEs as in Model III.



\end{document}